\pdfoutput=1 
\documentclass[a4paper,twoside,10pt,fleqn]{article}
\usepackage{helvet}

\usepackage[latin1]{inputenc}
\usepackage{amsfonts}
\usepackage{amsmath}
\usepackage{amssymb,marvosym,units}
\usepackage{amsthm}
\usepackage[cmtip,arrow, matrix, curve]{xy}
\usepackage{pb-diagram,pb-xy}
\usepackage{graphicx}
\usepackage{tensor}

\usepackage{hyperref}
\usepackage{zref, xkeyval, ifpdf, ifthen, calc, marginnote, pdfcomment}
\usepackage{bbm}

\hypersetup{
    pagebackref=true,
    bookmarks=true,        
    unicode=false,          
    pdftoolbar=true,        
    pdfmenubar=true,        
    pdffitwindow=true,     
    pdfstartview={FitH},    
    pdftitle={A holonomy groupoid formulation of Loop Quantum Gravity},   
    pdfauthor={Diana Kaminski},     
    pdfsubject={},   
    pdfkeywords={Loop Quantum Gravity, holonomy groupoid}, 
    pdfnewwindow=true,      
    colorlinks=false,      
    linkcolor=black,         
    pdfborder= 0 0 0,
}

\title{Algebras of Quantum Variables for Loop Quantum Gravity\\[5pt]
\textbf{VI. A holonomy groupoid formulation}}
\author{Diana Kaminski\\[3pt]
kaminski@math.uni-paderborn.de\\
\small{Europe - Germany}}
\date{August 19, 2011}

\newcommand{\A}{\begin{large}\mathcal{A}\end{large}}

\newcommand{\abroid}{\mathfrak{a}}

\newcommand{\Alg}{\begin{large}\mathfrak{A}\end{large}}

\newcommand{\Aut}{\begin{large}\mathfrak{Aut}\end{large}}

\newcommand{\CB}{\mathbb{C}}
\newcommand{\CD}{\mathcal{C}}

\newcommand{\g}{\mathfrak{g}}
\newcommand{\Goid}{\mathcal{G}}

\newcommand{\KD}{\mathcal{K}}

\newcommand{\la}{\langle}
\newcommand{\LD}{\mathcal{L}}

\newcommand{\mor}{\mathfrak{m}}

\newcommand{\PD}{\mathcal{P}}

\newcommand{\ra}{\rangle}

\newcommand{\ta}{\tilde\alpha}

\newcommand{\tvt}{\tilde\vartheta}

\newcommand{\FD}{\mathcal{F}}

\newcommand{\R}{\mathbb{R}}

\newcommand{\ZD}{\mathcal{Z}}

\newcommand{\ho}{\mathfrak{h}}
\newcommand{\Ho}{\mathfrak{H}}
\newcommand{\go}{\mathfrak{g}}

\newcommand{\ko}{\mathfrak{k}}

\DeclareMathOperator{\Ad}{Ad}

\DeclareMathOperator{\adj}{ad}

\DeclareMathOperator{\basic}{basic}

\DeclareMathOperator{\dif}{d}

\DeclareMathOperator{\Diff}{Diff}

\DeclareMathOperator{\Exp}{Exp}

\DeclareMathOperator{\thin}{\tiny{thin}}
\DeclareMathOperator{\hor}{hor}

\DeclareMathOperator{\Hol}{Hol}
\DeclareMathOperator{\Hom}{Hom}
\DeclareMathOperator{\Hoop}{Hoop}

\DeclareMathOperator{\id}{id}

\DeclareMathOperator{\Inti}{in}

\DeclareMathOperator{\Ima}{Im}
\DeclareMathOperator{\Iso}{Iso}

\DeclareMathOperator{\LG}{LG}

\DeclareMathOperator{\pr}{pr}

\newcommand{\gp}{{\gamma^\prime}}

\newcommand{\tg}{{\tilde\gamma}}
\newcommand{\tgp}{{\tilde\gamma^\prime}}

\newcommand{\gpp}{\gamma^{\prime\prime}}
\newcommand{\tgpp}{\tilde\gamma^{\prime\prime}}

\newcommand{\idf}{\mathbbm{1}}

\newcommand{\bra}{[}
\newcommand{\ket}{]}

\newcommand{\beq}{\begin{equation}\begin{aligned}}
\newcommand{\beqs}{\begin{equation*}\begin{aligned}}
\newcommand{\be}{\begin{flalign}}
\newcommand{\bes}{\begin{equation*}}
\newcommand{\eq}{\end{aligned}\end{equation}}
\newcommand{\eqs}{\end{aligned}\end{equation*}}
\newcommand{\ee}{\end{flalign}}
\newcommand{\ees}{\end{equation}}

\newtheorem{theo}{Theorem }[section]
\newtheorem{lem}[theo]{Lemma}
\newtheorem{rem}[theo]{Remark}
\newtheorem{prop}[theo]{Proposition}
\newtheorem{cor}[theo]{Corollary}

\newtheorem{defi}[theo]{Definition}

\newtheorem{Baxiom}{Barrett Axiom}

\newenvironment{proofs}[1][Proof ]{\noindent\textbf{#1}: }{\ \begin{flushright}
                                                                         \rule{0.5em}{0.5em}
                                                                        \end{flushright}}

\newcounter{exa}[section]

\newcounter{problem}[subsection]
 
\newcommand{\GGi}{\xymatrix{
  \Goid_1  \ar@<-2pt>[r] \ar@<2pt>[r] &  \Goid^0_1    \\
}}
\newcommand{\GGii}{\xymatrix{
  \Goid_2  \ar@<-2pt>[r] \ar@<2pt>[r] &  \Goid^0_2    \\
}}
\newcommand{\GGm}{\xymatrix{
  \Goid  \ar@<-1pt>[r]^{s} \ar@<1pt>[r]_{t} &  \Goid^0    \\
}}
\newcommand{\GGim}{\xymatrix{
  \Goid_1  \ar@<-1pt>[r]^{s_1} \ar@<1pt>[r]_{t_1} &  \Goid^0_1    \\
}}
\newcommand{\GGiim}{\xymatrix{
  \Goid_2  \ar@<-1pt>[r]^{s_2} \ar@<1pt>[r]_{t_2} &  \Goid^0_2    \\
}}

\newcommand{\PGm}{\xymatrix{
  \PD  \ar@<-1pt>[r]^{s} \ar@<1pt>[r]_{t} &  \Sigma    \\
}}

\newcommand{\PGs}{\PD\Sigma\rightrightarrows\Sigma}

\newcommand{\fPGm}{\xymatrix{
  \PD_\Gamma  \ar@<-1pt>[r]^{s} \ar@<1pt>[r]_{t} &  V_\Gamma    \\
}}
\newcommand{\PGsm}{\xymatrix{
  \PD\Sigma \ar@<-1pt>[r]^{s_{\PD\Sigma}} \ar@<1pt>[r]_{t_{\PD\Sigma}} &  \Sigma   \\
}}
\newcommand{\fPGms}{\xymatrix{
  \PD^s_\Gamma  \ar@<-1pt>[r]^{s} \ar@<1pt>[r]_{t} &  V_\Gamma    \\
}}

\newcommand{\fPSGm}{\xymatrix{
  \PD_\Gamma\Sigma  \ar@<-1pt>[r]^{s} \ar@<1pt>[r]_{t} &  V_\Gamma    \\
}}
\newcommand{\fPSG}{\xymatrix{
  \PD_\Gamma\Sigma  \ar@<-2pt>[r] \ar@<2pt>[r] &  V_\Gamma    \\
}}
\newcommand{\fgHGm}{\xymatrix{
  H(\Gamma)  \ar@<-1pt>[r]^/0.3em/{\hat s_H} \ar@<1pt>[r]_/0.3em/{\hat t_H} &  V_\Gamma    \\
}}

\newcommand{\fHGm}{\xymatrix{
  H_\Gamma  \ar@<-1pt>[r]^/0.3em/{\hat s_H} \ar@<1pt>[r]_/0.3em/{\hat t_H} &  V_\Gamma    \\
}}

\newcommand{\fGGm}{\xymatrix{
  \G^G_\Gamma  \ar@<-1pt>[r]^/0.3em/{s_P} \ar@<1pt>[r]_/0.3em/{t_P} &  V_\Gamma    \\
}}
\newcommand{\fGHm}{\xymatrix{
  \G^H_\Gamma  \ar@<-1pt>[r]^/0.3em/{s_P} \ar@<1pt>[r]_/0.3em/{t_P} &  V_\Gamma    \\
}}
\newcommand{\fG}{\frac{P\times P}{G}  \rightrightarrows \Sigma
}

\newcommand{\Group}{\Goid  \rightrightarrows \Goid^0
}

\newcounter{count}
\begin{document}
\maketitle
\begin{abstract}\noindent
The philosophy of the Loop Quantum Gravity approach is to construct the canonical variables by using the duality of infinitesimal connections and holonomies along loops. Based on this fundamental property for example the holonomy-flux $^*$-algebra \cite{LOST06} has been formulated. A generalisation of the one-to-one correspondence between infinitesimal objects: connections and curvature and path based objects: holonomy maps and parallel transports is used to replace the configuration space of the theory. This generalised duality is related to the concept of path connections and holonomy groupoids, which originally has been invented by Mackenzie \cite{Mack05} and which is presented shortly in this article. Finally these objects are used to propose some new algebras of quantum variables for Loop Quantum Gravity. 
\end{abstract}

\thispagestyle{plain}
\pdfbookmark[0]{\contentsname}{toc}
\tableofcontents

\section{Introduction}

In the previous articles \cite{Kaminski1,Kaminski2,Kaminski3,Kaminski4,KaminskiPHD} of the project \textit{AQV} different $^*$- or $C^*$-algebras for a theory of Loop Quantum Gravity have been presented, but none of these algebras contain a quantum analog of the classical variable curvature. A short overview about the ideas of the new holonomy groupoid formulation has been presented at the conference about ''Open problems in LQG'' \cite{Kaminskitalk}. It has been argued in \cite{Kaminski0} that, the Hamilton constraint cannot be quantised without a modification of this operator by replacing the curvature. The aim of this project about \textit{Algebras of Quantum Variables in LQG} is to find a suitable algebra of quantum variables of the theory. This algebra is specified by the fact that the quantum Hamilton constraint is an element of (or is affiliated with) this new algebra. Moreover the algebra is assumed to be generated by certain holonomies along paths, quantum fluxes and the quantum analogue of curvature.  

Quantisation of a gravitational theory in the context of LQG uses substantially the duality between infinitesimal objects and holonomies along paths in a path groupoid. Barrett \cite{Barrett91} has presented a roadmap for the construction of the configuration space of two physical examples: Yang-Mills and gravitational theories. He have suggested to consider all holonomy maps along loops in a certain loop group. In the first two subsection of section \ref{sec pathLiegroupoid} of this article different groups of loops and different groupoids of paths are presented. Then the holonomy maps of Barrett are further generalised for the example of a gauge theory. The construction is based on the concept of path connections in a Lie groupoid, which has been introduced by Mackenzie \cite{Mack05}. In section \ref{sec pathLiegroupoid} several examples for Lie groupoids and some of their properties are collected. The simpliest Lie groupoid is given by a Lie group $G$ over $\{e_G\}$. The new framework of Mackenzie allows to study instead of the holonomy maps, which have been presented by Barrett \cite{Barrett91}, generalised holonomy maps associated to path connections in a Lie groupoid. In general the holonomy map in a Lie groupoid is a groupoid morphism from a path groupoid to a Lie groupoid, which satisfy some new conditions. In \cite[Section 3.3.4.2]{KaminskiPHD} some new path groupoids, which are called path groupoids along germs, have been studied. For these certain path groupoids a general holonomy map in the groupoid $G$ over $\{e_G\}$ has been defined. This new holonomy map corresponds one-to-one to a smooth path connection. 
There exists another example of a Lie groupoid, which is given by the gauge groupoid associated to a principal fibre bundle. This groupoid is introduced in section \ref{sec pathLiegroupoid}. For gauge theories the holonomy maps in the gauge groupoid are defined as a groupoid morphisms from a path groupoid to the gauge groupoid.  These groupoid morphisms correspond uniquely to path connections, too. In this situation a path connection is an integrated infinitesimal smooth connection over a lifted path in the gauge groupoid w.r.t. a principal fibre bundle $P(\Sigma,G,\pi)$. Then the holonomy map is related to a parallel transport in the principal fibre bundle. Furthermore a holonomy map for a gauge theory defines a holonomy groupoid for a gauge theory. Notice that the ordinary holonomy map in the framework of LQG \cite[Section 2.2]{Kaminski1}, \cite[Section 3.3.4]{KaminskiPHD} has been defined by a groupoid morphism form the path groupoid to the simple Lie groupoid $G$ over $\{e_G\}$. Since in this case the holonomy mapping maps paths to the elements of the structure group $G$, the holonomy map does not define a parallel transport in $P(\Sigma,G)$.  
The duality between infinitesimal connections and these new holonomy maps are reviewed very briefly in section \ref{sec duality}. A detailed analysis why the theory of Mackenzie really generalises the examples of Barrett has been presented in \cite[Section 3.3.1 and 3.3.2]{KaminskiPHD}. Finally, the generalisation of Barrett's idea is to consider all general holonomy maps in a Lie groupoid as the configuration space of the theory. For example for gauge theories the set of holonomy maps in the gauge groupoid is defined in section \ref{subsubsec holmapdef}.

The next step is to find a replacement of the curvature. Moreover the particular decomposition of the Ashtekar connection and curvature associated to the Ashtekar connection and the way how the foliation is embedded into the spacetime $M$ is required to be encoded somehow in the quantum algebra of gravity. Clearly this is very hard to be obtained in a background independent manner. The problems of implementing infinitesimal structures like infinitesimal diffeomorphisms and curvature arise from the special choice of the analytic holonomy $C^*$-algebra \cite{Kaminski1}, \cite[Chapter 6]{KaminskiPHD}. Furthermore it turns out that the modification of the conditions for the holonomy mappings presented in \cite[Section 3.3.4]{KaminskiPHD} are not sufficient for the implementation of quantum curvature. This is the reason why the author suggests to replace the configuration space of the theory. Hence all holonomy maps for a gauge theory replace the original quantum configuration variables. Each holonomy map defines a holonomy groupoid such that a family of $C^*$-algebra depending on a holonomy groupoid associated to a path connection is constructable. The $C^*$-algebras are called holonomy groupoid $C^*$-algebras of a gauge theory and are presented in subsection \ref{subsec Calgebraholgroupoid}. In this framework it is used that,  instead of a measure on the configuration space a family of measures are defined on a Lie groupoid. This is more general than the original approach of a measure on the quantum configuration space in LQG. There the quantum configuration space of generalised connections restricted to a graph is identified with the product group $G^{\vert\Gamma\vert}$ of the structure group $G$. Finally for gauge theories the quantum curvature is implemented by the following theorem. The generalised Ambrose-Singer theorem, which has been given by Mackenzie \cite{Mack05}, states that the Lie algebroid of the holonomy groupoid of a path connection is the smallest Lie algebroid, which is generated from infinitesimal connections and curvature. Furthermore there is an action related to infinitesimal connections and curvature, since both objects are encoded as elements of a Lie algebroid, on the holonomy groupoid $C^*$-algebra. This is studied in section \ref{subsec Calgebraholgroupoid}, too.  
Summarising the algebra of quantum variables for a gauge theory is generated by the holonomy groupoid $C^*$-algebra of a gauge theory, the quantum flux operators and the Lie algebroid of the holonomy groupoid of a path connection. The construction of the algebra generated by the holonomy groupoid $C^*$-algebra of a gauge theory and the quantum flux operators is similarly to the holonomy-flux cross-product $C^*$-algebra \cite{Kaminski2,KaminskiPHD}. The development for groupoids has been presented by Masuda \cite{MasudaI,MasudaII}. In this article the ideas are illustrated in subsection \ref{subsec Calgebracrossgroupoid}. For the construction a left (or right) action of the holonomy groupoid on the $C^*$-algebra $C(G)$ of continuous functions on the Lie group $G$ is necessary to define a $C^*$-groupoid dynamical system. This object replaces the $C^*$-dynamical systems of $C^*$-algebras. The new cross-product algebra contains holonomies and in some appropriate sense this algebra is generated by curvature and fluxes.  A detailed description of this construction is a further project. 

Finally two important remarks are presented as follows. First recognize that, there exists a set of holonomy groupoids each associated to a path connection, which is contained in a set of path connections. This implies that, there is a set of holonomy groupoid $C^*$-algebras each associated to a path connection, too. Furthermore there are morphisms between these algebras. Hence there is a natural structure of a category available. 
Secondly the whole construction of the algebras basically depends on the chosen principal fibre-bundle $P(\Sigma,G,\pi)$ and consequently in particular on the base manifold $\Sigma$. Now the idea is to use the covariance principle, which has been developed by Brunetti, Fredenhagen and Verch \cite{BrunFredVerch01} in the framework of algebraic quantum field theory. Some first analysis is presented in section \ref{subsec background} and will be studied in a further work.  

The next step is to take into account that, the Hamiltonian framework of gravity is given w.r.t. the orthonormal framebundle on $\Sigma$. Therefore the gauge groupoid has to be replaced by the orthornormal frame groupoid $\FD\left( \frac{O^+(\Sigma,q)\times so(3)}{SO(3)}\right)$ over $\Sigma$. An analysis of this object has been presented by Mackenzie \cite{Mack05} and the implementation into the new holonomy groupoid formulation is a future project.

\section{The basic quantum operators}\label{chapter confspace}
\subsection{Path and Lie groupoids}\label{sec pathLiegroupoid}
The basic variables of LQG theory are derived from paths, graphs or groupoids on a smooth or analytic manifold $\Sigma$. The investigations start with the important work of Barrett \cite{Barrett91}, who has introduced the concepts of holonomy map and loop group (or thin holonomy group). For a construction of different classical variables it is worth to understand his ideas and how these concepts can be generalised. Indeed a more abstract theory has been developed by Mackenzie \cite{Mack05} independly from Barrett. In this article it is studied why the theory of Mackenzie replaces the concepts of Barrett. A short overview is given in the next paragraphs.

The smooth paths and loops defined in section \ref{subsec smoothloop} are the fundamental objects, which define holonomy groups. Furthermore a certain holonomy group called the fundamental group at a specific point in the manifold for smooth paths is illustrated. The group structure can be generalised to groupoids. Therefore the fundamental groupoid is introduced. The difference is the following. The fundamental group consists of homotopy classes of loops at a chosen point, whereas the fundamental groupoid is defined by the homotopy classes of paths between arbitrary points. In particular the fundamental groupoid is a Lie groupoid. For Lie groupoids new mathematical concepts are available, which have been introduced by Mackenzie. For example transformations in Lie groupoids are presented in section \ref{subsec LieGroupoidsact}. Another example for a Lie groupoid is the gauge groupoid, which is associated to a principal bundle and is studied in section \ref{subsec Liegroupoids} in detail. The LQG theory is basically a gauge theory, since the fundamental object is a principal fibre bundle. Therefore the gauge groupoid is used for a generalisation of the quantum variables given by the holonomy mappings. These generalised holonomy maps for a gauge theorys are introduced in section \ref{sec Groupmorphism}. The idea for these new objects is based on the duality of infinitesimal and integrated objects, which has been reformulated by Mackenzie and is presented very briefly in section \ref{sec duality}. The aim is to develop the basic framework to show in section \ref{sec duality} that Mackenzie's theory \cite{Mack05} about path connections and infinitesimal connections generalises the theory of Barrett for gauge theories.

\subsubsection*{Loop spaces, loop and holonomy group} \label{subsec smoothloop}

A curve in a smooth manifold $\Sigma$ is a (piecewise) smooth map $\gamma:I\rightarrow \Sigma$ where $I=[0,1]$. The basic objects are studied to derive the loop and holonomy group at a base point in the manifold $\Sigma$. 

First the composition of parametrized curves $\gamma_i:[0,1]\rightarrow\Sigma$ for $i=1,2$ is given by
\beq \gamma_1\circ\gamma_2=(\gamma_1\gamma_2)(t)=\left\{\begin{array}{ll} \gamma_1(2t) & \text{ for }t\in \text{[}0,\nicefrac{1}{2}\text{]}\\
\gamma_2(2t-1) & \text{ for }t\in \text{[}\nicefrac{1}{2},1\text{]}\end{array}
\right.\eq This relation is not an associative operation.
Therefore consider the following equivalence relation. Two curves $\gamma$ and $\gp$ are \hypertarget{rep-equiv}{\textbf{reparametrization equivalent}} iff there is an orientation preserving diffeomorphism $\phi: I\rightarrow I$ such that $\gp=\gamma\circ\phi$.

\begin{defi}
 A \textbf{loop} is a smooth continuous mapping of the unit interval $I=[0,1]$ into a (topological) space $\Sigma$ such that $\gamma(0)=\gamma(1)=v$. The collection of loops in $\Sigma$ with base point $v$ is called \textbf{loop space} $L\Sigma$.
\end{defi}

The quotient of the loop space at $v$ and reparametrization equivalence does not form a group. If additionally to reparametrization equivalence the algebraic relation  $\gamma\circ\gamma^{-1}\simeq\idf_v$ is required, then the loop space at $v$ modulo these equivalence relations forms a group. But there are several other homotopy equivalences, which implement such branch lines, available in literature. Another relation on the loop space $L\Sigma$ has been defined for example by Barrett \cite{Barrett91}. Two loops $\gamma,\gp$ are said to be thinly homotopic iff there exist a homotopy of the composed path $\gp^{-1}\circ\gamma$ to the trivial loop at $v$. The corresponding idea is that the loop $\gp^{-1}\circ\gamma$ shrinks to the trivial loop, which encloses no area. In other words, the definition of thin homotopy is given as follows.

\begin{defi}
A loop $\gamma$ is \hypertarget{thin-loop}{\textbf{thin}} iff there exists a smooth homotopy of $\gamma$ to the trivial loop with the image of the homotopy lying entirely within the image of $\gamma$, i.e. the homotopy
$\varrho:I\times I\rightarrow \Sigma$ for $s,t\in I$ satisfies $\varrho(1,t)=\gamma(t)$, $\varrho(0,t)=\idf_v$ and $\varrho(s,0)=\rho(s,1)=\idf_v$, where $\idf_v$ is the trivial loop at $v$, and for all $t,s\in I$ and $\Ima(\varrho)=\Ima(\gamma)$.
\end{defi}

Then Barrett has called two loops $\alpha,\beta$ thinly equivalent iff $\alpha\circ\beta^{-1}$ is thin. Caetano and Picken \cite{CaetPick94} have improved this definition.

\begin{defi}\cite[p.837]{CaetPick94}
Two loops $\gamma,\gp$ are said to be \textbf{thinly homotopic} iff there exists a finite sequence $\gamma_1,...,\gamma_n$ of loops such that $\gamma_1=\gamma$ and $\gamma_n=\gp$ and $\gamma_{i+1}^{-1}\circ\gamma_{i}$ is a thin loop for $i=1,..,n-1$. 
\end{defi}
Notice that, two loops only differing by a reparametrization are thinly homotopic.

\begin{lem}
The thin homotopy relation is an equivalence relation. 

Denote the equivalence class by $\{\gamma\}$ and the relation by $\sim_{\thin}$. 
\end{lem}

Observe that, for two thinly homotopic loops $\alpha$ and $\beta$ there exists a sequence \\$\alpha, \beta_2, \beta_3,...,\beta_{n-1},\beta$ such that
\beq \beta_2^{-1}\circ\alpha\circ\beta_3^{-1}\circ\beta_2 ...\circ\beta_{n-2}\circ\beta^{-1}\circ\beta_{n-1}\sim_{\thin}\idf_v\eq

Moroever there is another modification of homotopy equivalence given by Mackenzie \cite[p.218]{Mack05}. Mackenzie has called a restriction of a curve $\gamma$ on a subinterval of $I=[0,1]$ a \textbf{revision} of $\gamma$. Two curves $\gamma,\gp$ are called \textbf{equally good} iff $\gamma(1)=\gp(1)$ and either $\gamma$ is a revision of $\gp$ or, conversely, $\gp$ is a revision of $\gamma$. A \textbf{lasso} (for a cover $\{U_i\}$ of $\Sigma$) is a loop of the form $\gamma^{-1}\circ\beta\circ\gamma$ where the loop $\beta$ lies entirely in one neighborhood $U_i$.
\begin{defi}
A loop $\alpha$ is said to be \textbf{approximated at $v$} by finite product $\gp$ of lassos which is equally good as $\alpha$ iff each loop $\beta_i$ of the lasso $\gamma_i^{-1}\circ\beta_i\circ\gamma_i$ is contained in a neighborhood $U_i$ of $v$ and the product of lassos are homotopic to $\idf_v$.
\end{defi}
If $\alpha$ is (thinly) homotopic to a point, then from the smooth homotopy map $\varrho:I\times I\rightarrow \Sigma$ for $s,t\in I$ which satisfies $\varrho(1,t)=\gamma(t)$, $\varrho(0,t)=\idf_v$ and $\varrho(s,0)=\varrho(s,1)=\idf_v$ for all $t,s\in I$ and the finite sequence of appropriate loops, the product of lassos is constructed easily. Hence, a loop, which is homotopic to a point $v$, is approximated at $v$ by a suitable product of lassos, which is equally good as the loop.  

For the one-to-one correspondence between holonomy maps and horizontal lifts a further developed definition is useful. 
The following homotopy has been first introduced by Ceatano and Picken \cite{CaetPick94}. 
\begin{defi}\label{intimate homotopic1}\cite[section 4]{CaetPick94}
Two smooth loops $\gamma,\gp:I\rightarrow\Sigma$ are said to be \hypertarget{intimate-homotopy}{\textbf{intimate homotopic}} iff there exists a map $\varrho:I\times I\rightarrow \Sigma$ such that 
\begin{enumerate} 
\item $\varrho$ is smooth,
\item for every $s\in I$ it is true that $\varrho(s,t)\in L\Sigma$,
\item for $0\leq\epsilon\leq\nicefrac{1}{2}$ the map $\varrho$ satisfies
\beq 0\leq s\leq\epsilon, &\qquad \varrho(s,t)=\gamma(t)\\
 1-\epsilon\leq s\leq 1, &\qquad \varrho(s,t)=\gp(t)\\
 0\leq t\leq\epsilon, &\qquad \varrho(s,t)=\gamma(0)\\
 1-\epsilon\leq t\leq 1, &\qquad \varrho(s,t)=\gamma(1)\eq  and 
\item the rank of the differential $\dif \varrho(s,t)$ is smaller or equal $1$ for all $(s,t)\in [0,1]^2$. 
\end{enumerate}
\end{defi}

Ceatano and Picken have shown that, this relation is indeed an equivalence relation. The equivalence relation is denoted by $\sim_{\Inti}$. Moreover, this relation is a weakening of the thin homotopy relation. The intimate relation is stronger than the same-holonomy relation for all smooth connections introduced by Ashtekar and Isham \cite{AshIsh92}.

Now, for the different equivalence relations the quotient spaces can be considered. For example the quotient of the loop space $L\Sigma$ and intimate homotopic equivalence is a  group, which is called the \textbf{intimate fundamental group $\pi_1^{\Inti}(\Sigma,v)$ at base point $v$}. If the rank condition of the differential is omitted, then the quotient of $L\Sigma$ and the homotopy equivalence is the \textbf{fundamental group $\pi_1(\Sigma,v)$}. Finally the quotient of the loop space $L\Sigma$ and thinly homotopic equivalence is a group, too. This group is  called the \textbf{loop group $\LG(v)$ at $v$} or \textbf{thin fundamental group $\pi^1_1(\Sigma)$ at $v$}. 

Now, follow the ideas of Barrett presented in \cite{Barrett91}. He has required that, the configuration space is given by the set of certain mappings from the loop space at a base point $v$ to the structure group $G$ of a principal bundle such that these mappings arise as holonomy mappings. There is a set of conditions introduced by Barrett, which are called the Barrett axioms in this article.

\underline{\textbf{The Barrett Axioms}}
\begin{Baxiom}\label{Baxiom_group}\hypertarget{BAxiom1}{\textbf{(Group homomorphism)}} the map $\ho_A:\LG(v)\rightarrow G$ is a group homomorphism
\end{Baxiom} 
\begin{Baxiom}\label{Baxiom_rep}\hypertarget{BAxiom2}{\textbf{(Reparametrization invariance)}} for every orientation preserving diffeomorphism $\phi:I\rightarrow I$ the map $\ho_A$ satisfy
\[\ho_A(\gamma)=\ho_A(\gamma\circ\phi)\text{ for all }\gamma\in \LG(v)\]
\end{Baxiom} 
\begin{Baxiom}\label{Baxiom_thin}\hypertarget{BAxiom4}{\textbf{(Same-holonomy relation)}} for two thinly homotopic loops $\alpha$ and $\beta$ the maps are equal, i.e. \[\ho_A(\alpha)=\ho_A(\beta)\Leftrightarrow \alpha\sim_{\thin}\beta\]
\end{Baxiom} 
\begin{Baxiom}\label{Baxiom_smooth}\hypertarget{BAxiom3}{\textbf{(Smothness)}} for a smooth family $\{\gamma:U\rightarrow L\Sigma, U\text{ open subset of }\R^3\}$ of loops all compositions $\ho_A\circ\gamma:U\rightarrow G$ of the map $\ho_\Gamma$ with elements of this family is smooth.
\end{Baxiom}
The axioms \hyperlink{BAxiom2}{(BAxiom1)} and \hyperlink{BAxiom1}{(BAxiom2)} implement the algebraic structure. It follows that $\ho_A(\gamma^{-1})=\ho_A^{-1}(\gamma)$ and the value of two loops differing only by reparametrization are equivalent. The last axiom give rise to a topological structure on $G$. The smooth family of loops are maps $\gamma$ such that $\gamma:U\times\bra 0,1\ket\rightarrow LG(v)$ with $\gamma(v,t)=\breve\gamma(t)$ are smooth.
Moreover, every map $\ho_A$ which obeys \hyperlink{BAxiom3}{(BAxiom4)} is continuous. Therefore, a topology on $LG(v)$ inherited from \hyperlink{BAxiom3}{(BAxiom4)} is defined, which is called the Barrett topology for the loop group $LG(v)$. Summarising the last condition guarantees the differentiability of the bundle and lifting. A map $\ho_A:\LG(v)\rightarrow G$ is called \textbf{holonomy map} if the axioms: \hyperlink{BAxiom2}{(BAxiom1)}, \hyperlink{BAxiom1}{(BAxiom2)}, \hyperlink{BAxiom4}{(BAxiom3)} and \hyperlink{BAxiom3}{(BAxiom4)} are fulfilled. For a fixed holonomy map $\ho_A:LG\rightarrow G$ let the axioms: \hyperlink{BAxiom2}{(BAxiom1)}, \hyperlink{BAxiom1}{(BAxiom2)}, \hyperlink{BAxiom4}{(BAxiom3)} and \hyperlink{BAxiom3}{(BAxiom4)} be satisfied, then the \textbf{holonomy group $HG(v)$ at $v$} is defined by the set
\( \{\ho_A(\gamma):\gamma\in \LG(v)\} \).
The reconstruction theorem of Barrett \cite{Barrett91} explains the term holonomy map. 
\begin{theo}\label{Reconstruction Barrett}\cite[\textbf{Reconstruction theorem}]{Barrett91}\\
For a given connected manifold $\Sigma$ with base point $v$ and a holonomy map $\ho_A:L\Sigma\rightarrow G$, then there exists a principal bundle $P(\Sigma,G)$, a point in the fibre  $u\in \pi^{-1}(v)$, and a connection $A$ on $P$ such that $\ho_A$ is the holonomy map of the bundle. 
\end{theo}
In particular Barrett has given the mathematical background for treating the holonomy maps as the primary, and connections and curvature as derived objects of the theory.
\begin{theo}\label{Representation Barrett}\cite[\textbf{Representation theorem}]{Barrett91}\\ 
For a given connected, Hausdorff manifold $\Sigma$ there is a bijective correspondence between \begin{enumerate}                                                                                              \item a triple $(P,A,u)$ consisting of a principal $G$-bundle $P$, a connection $A$ on $P$ and a base point $u\in P$ and
\item a holonomy map \(\ho_A:L\Sigma(v)\rightarrow G\).                                                                                             \end{enumerate} 
\end{theo}

Therefore, it is straight forward to consider for a point $u$ in a principal bundle $P$, a connection $A$ on $P$ the set
\beq \Phi_u:=\{g\in G: \ho_A(\alpha)=ug\quad\forall\alpha\in \LG(v)\}
\eq whenever $\ho_A$ is a holonomy map. This set is a Lie subgroup of $G$ and it is equal to the holonomy group $HG(v)$, where $\pi(u)=v$. The constant loop $\idf_{v}:[0,1]\rightarrow v$ at $V$ defines the identity map $\ho_A(\idf_{v})=u$. 

Caetano and Picken \cite{CaetPick94} have modified the reconstruction and representation theorems \ref{Reconstruction Barrett} and \ref{Representation Barrett} by using the intimate homotopy equivalence relation \ref{intimate homotopic1}. They have improved the representation theorem by showing that from $\alpha\sim_{\Inti}\beta$, it follows that $\ho_A(\alpha)=\ho_A(\beta)$ and $\ho_A$ is a group homomorphism arising from a holonomy map defined as a horizontal lift associated to a smooth connection $A$. The complicated step was to show that intimate loops have the same holonomy. Furthermore, let $c:\pi_1^{\Inti}(\Sigma,v)\rightarrow \pi_1(\Sigma,v)$ be the canonical morphism and consider the group homomorphism
$ h: \pi_1(\Sigma,v)\rightarrow G$, which is connected to $\ho_A:\pi_1^{\Inti}(\Sigma,v)\rightarrow  G$ by $\ho_A:=h\circ c$. Then by the Ambrose-Singer theorem the class of holonomies, which are given by group homomorphisms $h$, are associated to flat connections. 

In general a purely algebraic equivalence relation is given as follows.
\begin{defi}\label{algebraic homotopy equivalence}
The quotient of the loop space $L\Sigma$ at $v$ by the \textbf{algebraic homotopy equivalence} refering to the relations 
\beq \gamma&\simeq\gamma\circ \idf_{v}\\
\gamma\circ\gamma^{-1}&\simeq\idf_{v}\\
(\gamma\circ\gp)\circ\gpp&\simeq\gamma\circ(\gp\circ\gpp)
\eq is called the \hypertarget{alg-loop-group}{\textbf{(algebraic) loop group $\LD(\Sigma,v)$}}, i.e.
where $\idf_{v}$ is the trivial loop in $L\Sigma$ at $v$.
\end{defi}
The most general holonomy mapping maps the (algebraic) loop group $\LD(\Sigma,v)$ to $G$. Furthermore Ashtekar and Isham \cite{AshIsh92}, Ashtekar and Lewandowski \cite{AshLew93} and Lewandowski \cite{Lew93} construct a group, which they call a \textbf{hoop group}. This group is the quotient of the loop space at $v$ modulo thin equivalence, reparametrisation equivalence and the same-holonomy relation for thinly homotopic loops and all holonomy mappings. 

\subsubsection*{Fundamental groupoids of path spaces}\label{subsec fundamental}

Now, the loops are generalised to paths. Consider a collection of (piecewise) smooth curves starting at a source point $v$ on a path connected manifold $\Sigma$. The collection of curves starting at $v$ is called the path space $P\Sigma^v$ at $v$. The set of all curves starting at all $v\in\Sigma$ is called the \textbf{path space $P\Sigma$}. The elements of the path space are called paths. A fibre of the path space $P\Sigma$ is the loop space $L\Sigma(v)$ at $v$. Let $s_{P\Sigma}:P\Sigma\rightarrow \Sigma$ and $t_{P\Sigma}:P\Sigma\rightarrow \Sigma$ be two surjective maps. 

There exists a generalisation of thinly homotopic and intimate homotopic equivalence on the path space, which lead to the definition of the thin and intimate fundamental groupoid. 

\begin{defi}
Two paths $\gamma,\gp:I\rightarrow P\Sigma$ such that $s(\gamma)=s(\gp)$, respectively, $t(\gamma)=t(\gp)$ are said to be \hypertarget{thin-path-homotopy}{\textbf{thin path-homotopic}} iff 
\begin{itemize}
  \item(\hypertarget{IH3}{Thin Path-Homotopic 1}) there exists a finite sequence $\gamma_1,...,\gamma_n$ of paths such that $\gamma_1=\gamma$ and $\gamma_n=\gp$ and $\gamma_{i+1}^{-1}\circ\gamma_{i}$ is a \hyperlink{thin-loop}{thin loop} for $i=1,..,n-1$. 
\end{itemize} 
Denote with $[\gamma]$ the equivalence class of thin path-homotopy. 

\label{thin path-homotopy}
 Two paths $\gamma,\gp:I\rightarrow P\Sigma$ such that $s(\gamma)=s(\gp)$, respectively, $t(\gamma)=t(\gp)$ are said to be \hypertarget{path-homotopy}{\textbf{intimate path-homotopic}} iff there is a map $\varrho:I\times I\rightarrow \Sigma$ such that
\begin{itemize}
 \item(\hypertarget{IH1}{Path-Homotopic 1}) $\varrho$ is (piecewise) smooth 
 \item(\hypertarget{IH2}{Path-Homotopic 2}) for $0\leq\epsilon\leq\nicefrac{1}{2}$ the map $\varrho$ satisfies
\beqs 0\leq s\leq\epsilon, &\qquad \varrho(s,t)=\gamma(t)\\
 1-\epsilon\leq s\leq 1, &\qquad \varrho(s,t)=\gp(t)\\
 0\leq t\leq\epsilon, &\qquad \varrho(s,t)=\gamma(0)\\
 1-\epsilon\leq t\leq 1, &\qquad \varrho(s,t)=\gamma(1)\eqs
 \item(\hypertarget{IH3}{Intimate Path-Homotopic 3}) the rank of the differential $\dif \varrho(s,t)$ is smaller or equal $1$ for all $(s,t)\in [0,1]^2$.
\end{itemize}

A map $\varrho:\bra 0,1\ket\times \bra 0,1\ket\rightarrow \Sigma$ satisfying (\hypertarget{IH1}{Path-Homotopic 1}), (\hypertarget{IH2}{Path-Homotopic 2}) and (\hypertarget{IH3}{Path-Homotopic 3}) is called a \textbf{smooth rank-one homotopy}.
\end{defi} 
The last condition \hyperlink{IH3}{(Path-Homotopic 3)} of intimate path-homotopy guarantees that the homotopy sweeps out a surface of vanishing area. The (smooth) \textbf{path-homotopy} equivalence (relativ to endpoints) is defined by \hyperlink{IH1}{(Path-Homotopic 1)} and \hyperlink{IH2}{(Path-Homotopic 2)}. Hence two paths, which differ only by a reparamtrization, are not path-homotopic equivalent. Recognize that intimate path-homotopy is stronger than (smooth) path-homotopy. Denote the intimate path-homotopic equivalence relation by $\sim_{\text{intimate path-hom.}}$.

Moreover, for two paths $[\gamma],[\gp]$ a composition operation $\cdot$ is defined if $\gamma(1)=\gp(0)$. 

\begin{defi}
The \hypertarget{thin-fundamental-groupoid}{\textbf{thin fundamental groupoid $\Pi^1_1(\Sigma,v)$ at $v$}} over $\Sigma$ is the quotient of the path space $P\Sigma^v$ at $v$ and thin path-homotopy equivalence. The source and target maps are $s_{\PD\Sigma}([\gamma])=s(\gamma)=v$ and $t_{\PD\Sigma}([\gamma])=t(\gamma)=\gamma(1)$, the constant path $\idf_v$ at $v$ give rise to the inclusion $v\mapsto\idf_v$, the multiplication is given by the concatenation 
\beq \begin{array}{rll}
 \text{[}\gamma\cdot\gp\text{]}(t)&=\gamma(2t)&\text{ for } 0\leq t\leq\nicefrac{1}{2}\\
\text{[}\gamma\cdot\gp\text{]}(t)&=\gp(2t-1)&\text{ for }\nicefrac{1}{2}\leq t\leq 1\end{array}\eq 
and the inverse element of a path is given by the reverse of the path $[\gamma]^{-1}=\gamma(1-t)$. 
\end{defi}

In the following omit the brackets $[\gamma]$ for elements of $\Pi^1_1(\Sigma,v)$. 
The vertex or loop group $\Pi^1_1\Sigma_v^v$ of $\Pi^1_1\Sigma$ at a base point $v$ is the thin fundamental group $\pi^1_1(\Sigma,v)$. 

The quotient of the path space $P\Sigma$ and the (piecewise) \hyperlink{intimate-homotopy}{intimate path-homotopic equivalence} given in definition \ref{thin path-homotopy} is called the  \textbf{intimate fundamental groupoid $\Pi_1^{\Inti}\Sigma$}  over $\Sigma$. The vertex group $\Pi_1^{\Inti}\Sigma_v^v$, which is given by all loops at $v$, of the groupoid $\Pi_1^{\Inti}\Sigma$ at a base point $v$ is the intimate fundamental group $\pi_1^{\Inti}(\Sigma,v)$. 

If $P\Sigma$ and the usual path-homotopy (relativ to the endpoints) equivalence is used, then the appropriate quotient is called the fundamental groupoid $\Pi_1\Sigma$ and the vertex group is the fundamental group $\pi_1(\Sigma,v)$. If additionally $\Sigma$ is connected, then the fibres $\Pi_1\Sigma^v$ are the universal covering spaces of $\Sigma$. Remark that the fundamental group does not coincide with the loop group defined by thin homotopy in general. But the thin fundamental group is a quotient of the fundamental group, since, thin homotopy is a restricted notion of the usual homotopy equivalence on loop spaces. 
\begin{rem}\label{rem funda}
The \textbf{fundamental groupoid} over $\Sigma$ is presented by the set
\beqs \Pi_1(\Sigma):=\{ (v,[\gamma],w):v,w\in\Sigma,[\gamma]\text{ path-homotopy class of paths in }\PD\Sigma\text{ s.t. }\gamma(0)=v,\gamma(1)=w\}
\eqs equipped with quotient topology of the compact open topology on $\PD\Sigma$.

The map $s_{\Pi_1\Sigma}\times t_{\Pi_1\Sigma}:\Pi_1\Sigma\rightarrow \Sigma\times\Sigma$ is the covering map. 
\end{rem}
In general consider the path space modulo reparametrisation equivalence. Denote the this quotient also by $\PD\Sigma$. Then the following groupoid can be constructed.

\begin{defi}\label{algebraic path groupoid} A (algebraic) \textbf{path groupoid} $\PD\Sigma$ over $\Sigma$ is a pair  $(\PD\Sigma, \Sigma)$ equipped with the following structures: 
\begin{enumerate} 
\item two surjective maps \(s_{\PD\Sigma},t_{\PD\Sigma}:\PD\Sigma\rightarrow \Sigma\) called the source and target map,
\item the set \(\PD\Sigma^2:=\{ (\gamma_i,\gamma_j)\in\PD\Sigma\times\PD\Sigma: t(\gamma_i)=s(\gamma_j)\}\) of composable pairs of paths,
\item the  composition \(\circ :\PD\Sigma^2\rightarrow \PD\Sigma,\text{ where }(\gamma_i,\gamma_j)\mapsto \gamma_i\circ \gamma_j\),
\item the inversion \(\gamma_i\mapsto \gamma_i^{-1}\) of a path $\gamma_i$
\item object inclusion map \(\iota:\Sigma\rightarrow\PD\Sigma\) and
\item $\PD\Sigma$ modulo the algebraic equivalence relations generated by
\beqs \gamma_i^{-1}\circ \gamma_i\simeq \idf_{s(\gamma_i)}\text{ and }\gamma_i\circ \gamma_i^{-1}\simeq \idf_{t(\gamma_i)}
\eqs 
\end{enumerate}
Shortly, write $\PGsm$. 
\end{defi} 

\subsubsection*{General Lie and gauge groupoids}\label{subsec Liegroupoids}

For smooth paths Lie groupoids are naturally available. The advantage of Lie groupoids is the familiarity to Lie groups. In particular for Lie groupoids measures exists.
Hence in this section a short mathematical overview about important structures of Lie groupoids are collected. For a detail study refer to Mackenzie \cite{Mack05}.

\begin{defi}
A \textbf{Lie (or smooth) groupoid $\Goid$ over $\Goid^0$} is a groupoid where $\Goid$ and $\Goid^0$ are smooth manifolds (additionally, $\Goid^0$ and $\Goid^v$ for all $v\in\Goid^0$ are Hausdorff), $s_\Goid$ and $t_\Goid$ are smooth surjective submersions such that $\Goid^{(2)}$ is a smooth submanifold of the product manifold $\Goid\times\Goid$, the inclusion $i:\Goid^0\rightarrow\Goid$, the multiplication $\cdot$ and the inversion are smooth maps.  

A groupoid $\Goid$ is \textbf{transitive} if for each pair $(v,w)\in\Goid^0\times\Goid^0$ there is a morphism $\gamma\in\Goid$ such that $s_\Goid(\gamma)=v$ and $t_\Goid(\gamma)=w$. A Lie groupoid $\Goid$ is \hypertarget{loc-triv-groupoid}{\textbf{locally trivial}} if the map $s_\Goid\times t_\Goid:\Goid\rightarrow\Goid^0\times\Goid^0$, called anchor of $\Goid$, is a surjective submersion.
\end{defi}

In particular each locally trivial Lie groupoid is transitive. A vector bundle $E\overset{q}{\rightarrow}\Sigma$, which admits a group bundle structure, since, $s_E=t_E=q$ and the composition is fibrewise addition $E_x\times E_x\rightarrow E_x$, is a Lie groupoid $E$ over $\Sigma$ such that this groupoid is not locally trivial.

\begin{defi}Let $\Goid$ be a Lie groupoid on $\Sigma$.

A \textbf{Lie subgroupoid} of $\Goid$ is a Lie groupoid $\Goid^\prime$ on $\Sigma^\prime$ together with injective immersions $\iota:\Goid^\prime\rightarrow\Goid$ and $\iota_0:\Sigma^\prime\rightarrow\Sigma$ such that $(\iota,\iota_0)$ is a morphism of Lie groupoids. 
\end{defi}

In this article it is assumed that, the fibres $\Goid^u$ and $\Goid_u$ are connected for all $u\in\Goid^0$.

\begin{defi}
A \hypertarget{Lie groupoid-morphism}{\textbf{Lie groupoid morphism}} between two Lie groupoids $\FD$ over $\FD^{0}$ and $\Goid$ over $\Goid^{0}$ consists of two smooth maps  $\ho:\FD\rightarrow\Goid$  and $h:\FD^0\rightarrow\Goid^0$ such that
\beqs (\hypertarget{G1}{G1})\qquad \ho(\gamma\circ\gp)&= \ho(\gamma)\ho(\gp)\text{ for all }(\gamma,\gp)\in \FD^{(2)}\eqs
\beqs (\hypertarget{G2}{G2})\qquad s_\Goid(\ho(\gamma))&=h(s_{\FD}(\gamma)),\quad t_\Goid(\ho(\gamma))=h(t_{\FD}(\gamma))\eqs 
\end{defi}

\begin{defi}
A morphism $\ho:\FD\rightarrow\Goid$ and $h:\FD^0\rightarrow\Goid^0$ is an \textbf{isomorphism of Lie groupoids} if $\ho$ and $h$ are diffeomorphisms. 
\end{defi}

The simpliest example for a Lie groupoid is given by the gauge groupoid.

\begin{defi}
Let $G$ act on the right of the product $P\times P$ of a principal bundle $P:=P(\Sigma,G)$ by
\beqs g(u,p)=(ug,pg)\eqs and denote the orbit of $(u,p)$ by $\langle u,p\rangle$ and the set of orbits $\frac{P\times P}{G}$. Then $\Goid$ denote the groupoid $\fG$, called the \hypertarget{gauge-groupoid}{\textbf{gauge groupoid associated to $P(\Sigma,G)$ with base $\Sigma$}}, with
\beqs &s_P(\langle u,p\rangle):=(\pi\circ\pr_1)(\langle u,p\rangle)=\pi(u),\quad t_P(\langle u,p\rangle):=(\pi\circ\pr_2)(\langle u,p\rangle)=\pi(p)\\
&i(v)=\idf_v:=\langle u,u\rangle\text{ where } u\ni\pi^{-1}(v)\eqs
\beqs
\langle u_1,p_1\rangle\cdot\langle u_2,p_2\rangle: = \langle u_1\delta(p_1,u_2) ,p_2\rangle &\text{ for all }\langle u_i,p_i\rangle=:e_i, i=1,2\\
&\text{ such that } t_P(e_1)=s_P(e_2)\\
&\text{ where }\delta:P\times P\rightarrow G,\delta(u,ug)=g
\eqs such that the groupoid multiplication is smooth, the source map $s_P:\Goid\rightarrow \Sigma$ is a surjective submersion.

The inverse  is given by
\beqs \la u,p\ra^{-1}:=\la p,u\ra\eqs
\end{defi}

Set $\delta: G\times G\rightarrow G$ to be the difference map which is defined by $\delta(g,h):=g^{-1}h$. Then compute
\beq \la u,p\ra\la u,p\ra^{-1}=\la u,p\ra\la p,u\ra =\la u\delta(p,p),u\ra=\idf_v\eq and
\beq \la u,p\ra\la u,p\ra =\la u\delta(p,u), p\ra\eq
Set $p = ux$ and $u^\prime=py$ and derive
\beq \la u,p\ra\la u,p\ra \la p,u^\prime\ra\la p,u^\prime\ra& =\la u\delta(p,u), p\ra\la p\delta(u^\prime,p), u^\prime\ra=\la u x, p\ra\la p y, u^\prime\ra =\la p\delta(p,py),u^\prime\ra\\& =\la py,u^\prime\ra=\idf_v\eq

\begin{lem} The map
\beqs H: P\times P\rightarrow \Goid, h:P\rightarrow\Sigma\eqs 
is a Lie groupoid homomorphism from the pair groupoid $P\times P$ over $P$.
\end{lem}

\begin{cor}
The gauge groupoid $\Goid$ of a principal bundle $P(\Sigma,G,\pi)$ is a locally trivial Lie groupoid.
\end{cor}

Another Lie groupoid is the fundamental groupoid $\Pi_1\Sigma$ over $\Sigma$, which has been introduced in remark \ref{rem funda}. There is a correspondence between this fundamental groupoid and a certain gauge groupoid.
\begin{cor}
The fundamental groupoid $\Pi_1\Sigma$ is the gauge groupoid of the principal bundle $\tilde \Sigma( \Sigma,\pi(\Sigma))$ where $\tilde\Sigma$ is the universal cover of $\Sigma$.
\end{cor}

\begin{lem}Let $\Sigma$ be a connected manifold.

Then the Lie groupoid $\Pi_1\Sigma$ is connected. 
\end{lem}

For the generalisation of the concept of Barrett's duality betwen smooth connections and holonomy maps the following objects are necessary. The full detailed mathematics can be found in the book of Mackenzie \cite{Mack05}.

\begin{prop}
Let $\Goid$ be a locally trivial Lie groupoid. Then
\begin{enumerate}
 \item the isotropy groups $\Goid^u_u$ of $\Iso(\Goid)=\{\Goid^u_u\}_{u\in\Goid^0}$ are isomorphic to Lie groups.
 \item For each $u\in\Goid^0$ the fibre $\Goid^u:=s_\Goid^{-1}(u)$ is a differentiable principal bundle over $\Goid^0$ with the surjection $t_\Goid$, a smooth and free left action $L:\Goid^u_u\times\Goid^u\rightarrow\Goid^u$ and the isotropy group $\Goid^u_u$ as structure group. $\Goid^u(\Goid^0,\Goid^u_u,t_\Goid)$ is called \textbf{vertex bundle at $u$}.
 \item for $u,v\in\Goid^0$ and an element $\theta\in \Goid^u_v$ there is an isomorphism of principal bundles over $\Goid^0$
\beqs L_{\theta}(\id_{\Goid^0},I_\theta):\Goid^v(\Goid^0,\Goid^v_v,t_\Goid)\rightarrow\Goid^u(\Goid^0,\Goid^u_u,t_\Goid)\\
\text{ where }I_\theta:\Goid^v_v\rightarrow\Goid_u^u,\quad \gamma\mapsto \theta\circ\gamma\circ\theta^{-1}\eqs
\end{enumerate}
\end{prop}

\begin{cor}\label{isomorphismAsso} Let $\Goid$ be a locally trivial Lie groupoid over $\Goid^0$.
 
Then the map 
\beq \frac{\Goid^u\times\Goid^u}{\Goid^u_u}\rightarrow\Goid,\quad \langle\tgp,\tg\rangle\mapsto \tgp\circ\tg^{-1}\eq
from the gauge groupoid of the vertex bundle at $u$ to the Lie groupoid is an isomorphism of Lie groupoids over $\Goid^0$.

The map
\beq \frac{\Goid^u\times\Goid^u_u}{\Goid^u_u}\rightarrow\Iso(\Goid),\quad <\tg,\alpha> \mapsto \tg \circ \alpha\circ\tg^{-1}\eq
from the associated fibre bundle to the vertex bundle at $u$ w.r.t. the action of inner-translation on $\Goid^u_u$ to the isotropy Lie group bundle is an isomorphism of Lie groupoids over $\Goid^0$.
\end{cor}

Let $\Goid$ be a locally trivial Lie groupoid with connected fibres $\Goid^v$. Denote $\PD^{s_\Goid}(\Goid)$ be the set of continuous and piecewise-smooth paths $\tg:I\rightarrow\Goid$ (where $I=[0,1]$) for which $s_\Goid \circ\tg:\text{[}0,t\text{]}\rightarrow \Goid^0$ is constant for all elements of $\tg\in\PD^{s_\Goid}(\Goid)$ and $t\in I$. 

\begin{defi}\label{psmoothhomo1}
Two paths $\gamma$ and $\gp$ in $\PD^{s_\Goid}(\Goid)$ are called \hypertarget{psmoothhomo}{\textbf{$s_\Goid$-homotopic}} $\gamma\overset{s_\Goid}{\sim}\gp$ , if $s_\Goid(\gamma)=s_\Goid(\gp)$, respectively, $t_\Goid(\gamma)=t_\Goid(\gp)$ and there is a continuous and piecewise-smooth map $\varrho:I\times I\rightarrow \Goid$ such that (\hyperlink{IH2}{Path-Homotopic 2}) property:
\beqs 0\leq s\leq\epsilon, &\qquad \varrho(s,t)=\gp(t)\\
 1-\epsilon\leq s\leq 1, &\qquad \varrho(s,t)=\gamma(t)\\
 0\leq t\leq\epsilon, &\qquad \varrho(s,t)=\gamma(0)\\
 1-\epsilon\leq t\leq 1, &\qquad \varrho(s,t)=\gamma(1)\\
\eqs is satisfied and for each $s\in I$ the map $\varrho(s,t)$ is an element of $\PD^{s_\Goid}(\Goid)$. The equivalence class is denoted by $[\tg]$ for all $\tg\in\PD^{s_\Goid}(\Goid)$. 
\end{defi}

Finally in general for every Lie groupoid $\Goid$ there exists an associated Lie algebroid $A\Goid$. 
\begin{defi}
A \textbf{Lie algebroid $A\Goid$ associated to a transitive Lie groupoid $\Goid$} is a vector bundle over $\Goid^0$, which is equipped with a vector bundle map $a: A\Goid\rightarrow T\Goid^0$,  which is called anchor, a Lie bracket $\bra.,.\ket_{A\Goid}$ on the space $\Gamma(A\Goid)$ of smooth sections of $A\Goid$, satisfying the following compatibility conditions 
\begin{enumerate}
 \item $\R$-bilinear
 \item alternating and Jacobi identity
 \item $\bra X, fY\ket=f\bra X,Y\ket + a(X)(f)Y$ for all $X,Y\in\Gamma(A\Goid)$ and $f\in C^\infty(\Sigma)$
\item $a(\bra X,Y\ket)=\bra a(X),a(Y)\ket$ for all $X,Y\in\Gamma(A\Goid)$ and $f\in C^\infty(\Sigma)$.
\end{enumerate}\end{defi}
In particular the vector bundle $T_v(\Goid)$ over $\Goid^0$ is a Lie algebroid.

\subsubsection*{Transformations in a Lie groupoid}\label{subsec LieGroupoidsact}
After the definitions of the basic objects some transformations are introduced. The definitions are borrowed from Mackenzie \cite{Mack05}. 
\begin{defi}\label{defi transfLie} Let $\Goid$ be a Lie groupoid on the base $\Goid^0$, for $\tg\in\Goid$ with $s_\Goid(\tg)=v$ and $t_\Goid(\tg)=w$ 
the \textbf{left-translation corresponding to $\Goid$} is defined by
\beqs L_{\tg}:\Goid^{w}\rightarrow \Goid^{v},\quad \tvt\mapsto \tg\circ \tvt \eqs
and the \textbf{right-translation corresponding to $\Goid$}
\beqs R_{\tg}:\Goid_{v}\rightarrow \Goid_{w},\quad \tvt\mapsto \tvt\circ\tg \eqs 
\end{defi}

\begin{defi}
\hypertarget{left-translation in a Lie groupoid}{\textbf{A left-translation in a Lie groupoid $\Goid$}} over $\Sigma$ is a pair of diffeomorphisms 
\beqs\Phi:\Goid\rightarrow\Goid\text{ and }\varphi:\Sigma\rightarrow\Sigma\eqs 
such that 
\beqs s_\Goid(\Phi(\gamma))=\varphi(s_\Goid(\gamma)),\quad t_\Goid(\Phi(\gamma))=\varphi(t_\Goid(\gamma))\text{ for all }\gamma\in\Goid\eqs
and, moreover, 
\beqs \Phi^v:\Goid^v\rightarrow \Goid^{\varphi(v)},\quad \gamma\mapsto L_{\theta}(\gamma)\text{ for some }\theta\in\Goid_{v}^{\varphi(v)}\text{ and all }\gamma\in\Goid^v\eqs

A \textbf{global bisection of $\Goid$} is a smooth map $\sigma:\Sigma\rightarrow\Goid$ which is right-inverse to $s_\Goid:\Goid\rightarrow\Sigma$ (w.o.w. $s_\Goid\circ\sigma=\id_\Sigma$) and such that $t_\Goid\circ\sigma:\Sigma\rightarrow\Sigma$ is a diffeomorphism. The set of bisections on $\Goid$ is denoted $\mathfrak{B}(\Goid)$. Denote the image of the bisection by $L:=\{\sigma(v):v\in\Sigma\}$ which is a closed embedded submanifold of $\Goid$.
\end{defi}

The set of global bisections $\mathfrak{B}(\Goid)$ forms a group, where the multiplication is given by
\beqs (\sigma\ast\sigma^\prime)(v): =\sigma^\prime(v)\circ\sigma(t_\Goid(\sigma^\prime(v)))\text{ for }v\in\Sigma,\eqs
the object inclusion $v\mapsto\idf_v$ of $\Goid$, where $\id_v$ is the unit morphism at $v$ in $\Goid^v_v$, and the inversion is given by
\beqs \sigma^{-1}(v):=\sigma((t_\Goid\circ\sigma)^{-1}(v))^{-1}\text{ for }v\in \Sigma\eqs

Define for a given bisection $\sigma$, the right-translation in $\Goid$ by the map
\beq R_\sigma:\Goid\rightarrow\Goid, \quad \gamma\mapsto \gamma\circ\sigma(t_\Goid(\gamma)) \eq

The map $\sigma\mapsto R_\sigma$ is a group isomorphism, i.e. $R_{\sigma\ast\sigma^\prime}=R_\sigma\circ R_{\sigma^\prime}$. Whereas $\sigma\mapsto t_\Goid\circ\sigma$ is a group morphism from $\mathfrak{B}(\Goid)$ to the group of diffeomorphisms $\Diff(\Sigma)$.

Notice that local bisection are defined to be maps $\sigma:U\rightarrow \Goid$ where $U$ is an open subset in $\Sigma$.

\begin{defi}Let $\Goid$ be a Lie groupoid on $\Sigma$ , and fix a bisection $\sigma\in\mathfrak{B}(\Goid)$.

Then define the \textbf{left-translation} 
\beqs L_\sigma:\Goid\rightarrow\Goid,\quad \gamma\mapsto \sigma((t_\Goid\circ\sigma)^{-1}(s_\Goid(\gamma)))\circ\gamma\eqs
and the \textbf{inner-translation} is given by
\beqs I_\sigma:\Goid\rightarrow\Goid,\quad \gamma\mapsto \sigma(s_\Goid(\gamma))^{-1}\circ\gamma\circ\sigma(t_\Goid(\gamma))\eqs
which is an isomorphism of Lie groupoids over $t_\Goid\circ\sigma:\Sigma\rightarrow\Sigma$.
\end{defi}

Clearly $L_{\sigma\ast\sigma^\prime}=L_\sigma\circ L_{\sigma^\prime}$ and $I_{\sigma\ast\sigma^\prime}=I_\sigma\circ I_{\sigma^\prime}$. $L_{\sigma^{-1}}(\gamma)=\sigma(s_\Goid(\gamma))^{-1}\circ\gamma$ and $I_\sigma=R_\sigma\circ L_{\sigma^{-1}}=L_{\sigma^{-1}}\circ R_\sigma$ yield.

\subsubsection*{Transformations in the gauge groupoid associated to a principal fibre bundle $P(\Sigma,G,\pi)$:}
\begin{lem}
Consider the automorphisms $\varphi:P\rightarrow P$ on the principal bundle $P(\Sigma,G,\pi)$, a diffeomorphism $\varphi_0$ on $\Sigma$ and the identity map $\id$ on the structure group $G$.
Assume $\pi\circ\varphi=\varphi_0\circ\pi$, $\varphi(ug)=\varphi(u)g$ for all $u\in P$ and $g\in G$. 

For a fixed element $u\in P$ such that $\pi^{-1}(u)=v$ yields, set 
\beqs \sigma(v):=\langle u,\varphi(u)\rangle\eqs
Then $\sigma$ is smooth, since $\pi$ is an surjective submersion and $\sigma$ is a bisection of the gauge groupoid $\fG$. 

Moreover, there exists an action $I_\sigma: \frac{P\times P}{G}\longrightarrow \frac{P\times P}{G}$ given by
\beqs I_\sigma(\langle u,p\rangle):=\langle\varphi(u),\varphi(p)\rangle\eqs
The automorphism $\varphi(\varphi_0,\id)$ is called \textbf{gauge and diffeomorphism transformation} in $P(\Sigma,G,\pi)$.
\end{lem}

\subsection{Duality of connections and holonomies}\label{sec duality}

A generalisation of Barrett's duality \cite{Barrett91} is given by the duality of infinitesimal connections of a principal bundle $P(\Sigma,\pi,G)$ and path connections in a corresponding Lie groupoid. In this particular case the Lie groupoid is given by the gauge groupoid. The theory of this duality in a more general framework (of transitive Lie groupoids) has been invented by Mackenzie \cite{Mack05}. In this section a very short overview about the basic structures is given. The infinitesimal geometric objects are presented for the special case of a gauge theory. Moreover these definitions coincide with the objects, which are usually given in books about differential geometry.  The duality of infinitesimal connections and holonomies is analysed in the more general context of Mackenzie. The idea is to use the more enhanced framework for a definition of quantum variables for a gauge theory and hence new algebras. 

\subsubsection{Infinitesimal geometric objects for a gauge theory}\label{subsec infobj}
In this section the basic geometric objects and their relations in the framework of Mackenzie are collected. For a study refer to the book \cite{Mack05} of Mackenzie.

Let $P(\Sigma,G,\pi)$ be a principal bundle where $P$ is a smooth manifold. The object $T\Sigma$ over $\Sigma$ defines a Lie algebroid.
The vector bundle morphism $\frac{TP}{G}\overset{\pi_\ast}{\rightarrow}T\Sigma$ constructed from the principal $G$- bundle $P\overset{\pi}{\rightarrow}\Sigma$ defines a Lie algebroid structure on $\frac{TP}{G}$. This follows from the fact that, there is a commutator map $[.,.]$ defined on the sections $\Gamma(T\Sigma)$ which is transfered to a bracket $\bra .,.\ket_P$ on  $\frac{TP}{G}$. Hence $\frac{TP}{G}$ over $\Sigma$ is a Lie algebroid.
The object $\Goid:=\frac{P\times P}{G}\rightrightarrows\Sigma$ is a transitive Lie groupoid, which is called the \textbf{gauge groupoid} of a principal bundle $P(\Sigma,G,\pi)$.

On the one hand there exists an exact sequence of Lie groupoids 
\beq\label{exactsequofLG} \frac{P\times G}{G}\overset{\iota}{\rightarrowtail} \frac{P\times P}{G}\overset{\tilde\pi}{\twoheadrightarrow} \Sigma\times\Sigma\eq where $\frac{P\times G}{G}$ is a Lie group bundle and $\Sigma\times\Sigma\rightrightarrows\Sigma$ is the \textbf{pair groupoid}. The maps $\iota,\tilde\pi$ are groupoid morphisms, $\iota$ is an embedding, $\tilde\pi$ is a surjective submersion and $\Ima(\iota)=\ker(\tilde\pi)$. Notice that, $\tilde\pi$ is given by the map $s_P\times t_P$, where $s_P,t_P$ are the source and target map of the Lie groupoid $\fG$. 

On the other hand there is an exact sequence of the Lie algebroids $\frac{TP}{G}$ and $T\Sigma$ over $\Sigma$, called the \textbf{Atiyah sequence}, which is given by
\beq\label{exactsequofAG}\frac{P\times \g}{G}\overset{j}{\rightarrowtail} \frac{TP}{G}\overset{\pi_\ast}{\twoheadrightarrow} T\Sigma\eq
such that $j,\pi_\ast$ are vector bundle morphisms. The Lie algebroid bundle $\frac{P\times \g}{G}=\ker(\pi_\ast)$ is called the \textbf{adjoint bundle}.

A \textbf{Lie algebroid connection} $\gamma_A$ is a right splitting of the exact sequence \eqref{exactsequofAG}, which is a map
$\gamma_A:T\Sigma\rightarrow \frac{TP}{G}$ such that $\pi_\ast\circ\gamma_A=\id_{T\Sigma}$. A Lie algebroid connection $\gamma_A$ is also called an \textbf{infinitesimal connection} and is shortly denoted by $A$. The formulation as a right splitting of an exact sequence has the advantage that this definition simply generalises to transitive Lie groupoids and transitive Lie algebroids. This has been derived in \cite[Section 3.5, Section 5.2]{Mack05} by Mackenzie. In general an \textbf{infinitesimal connection in a transitive Lie groupoid} $\Goid$ over $\Sigma$ is a morphism of vector bundles $\gamma:T\Sigma \rightarrow A\Goid$ over $\Sigma$ such that $a\circ\gamma=\id_{T\Sigma}$ and $a$ is the anchor of the Lie algebroid $A\Goid$ associated to a transitive Lie groupoid $\Goid$. Hence, in the context of a gauge theory the infinitesimal connections are also called the infinitesimal connections in the gauge groupoid. The adjoint connection $\nabla^{\adj}:=\adj\circ \gamma_A$ is defined as the commutator $j(\nabla^{\adj}_X(V))=\bra \gamma_A X,j(V)\ket$ for $V\in\Gamma\big(\frac{P\times \go}{G}\big)$ and $X$ is a smooth vector field $\mathfrak{X}(\Sigma)$.

Notice the following structure. There exists a map $l:T\Sigma\longrightarrow \frac{P\times\go}{G}$ such that the Lie algebroid connection $\gamma_{A^\prime}$ decomposes into a sum $\gamma_{A^\prime}:=\gamma_A + j\circ l$, where $\gamma_A$ is another Lie algebroid connection.

A \textbf{connection reform} $\omega:TP\rightarrow P\times \g$ which is $G$-equivariant and horizontal, w.o.w. $\omega^{/G}\in\Omega^1_{\basic}(P,\g)^G$, corresponds to a morphism $ \omega^{/G}:\frac{TP}{G}\longrightarrow\frac{P\times\go}{G}$ of vector bundles over $\Sigma$ such that $\omega^{/G}\circ j=\id_L$ where $L=\frac{P\times \g}{G}$.
There is a bijective correspondence betwen the connection $\gamma_A$ and a connection reform $\omega$,  such that
$j\circ\omega^{/G}+ \gamma_A\circ \pi_\ast=\id_{A\Goid}$ where $A\Goid:=\frac {TP}{G}$.

The \textbf{curvature} (for a gauge theory) is a skew-symmetric vector bundle map $\bar R_{A}:T\Sigma\oplus T\Sigma\rightarrow \frac{P\times\g}{G}$ such that $j(\bar R_A(X,Y))=\gamma_A([X,Y])-[\gamma_A(X),\gamma_A(Y)]$ for $X,Y\in \mathfrak{X}(\Sigma)$ being smooth sections in $T\Sigma$. 

\subsubsection{Integrated infinitesimals, path connections, holonomy groupoids and holonomy maps in Lie groupoids}\label{subsec pathconnection}

The concept of integrated infinitesimal connections over a lifted path in a Lie groupoid $\Goid$ over $\Sigma$, which is called a path connection $\Lambda$ on $\Goid$, has been developed by Mackenzie \cite{Mack05}. In this section this more general framework of a path connection in a Lie groupoid is used. The main object derived from this path connection is the following: a Lie algebroid connection $\gamma_A$ in a Lie algebroid $A\Goid$ of a locally trivial Lie groupoid $\Goid$ over a connected smooth base manifold $\Sigma$. The Lie algebroid connection give rise to a notion of lifting paths in $\Sigma$ to paths in a Lie groupoid $\Goid$ over $\Goid^0$. Moreover, associated to a path connection there exists the holonomy groupoid and the holonomy map for a given Lie groupoid. Note that, in the next section the fundamental definitions and theorems are collected and are generalised to groupoids. For a detailed study refer to Mackenzie \cite{Mack05}.

\paragraph*{Path connection and holonomy maps for a groupoid\\[10pt]}

Let $\Goid$ be a locally trivial Lie groupoid with connected fibres $\Goid^v$. Then recall the set $\PD^{s_\Goid}(\Goid)$ of continuous and piecewise-smooth paths $\tg:I\rightarrow\Goid$ (where $I=[0,1]$) for which $s_\Goid \circ\tg:\text{[}0,t\text{]}\rightarrow \Goid^0$ is constant for all elements of $\tg\in\PD^{s_\Goid}(\Goid)$ and $t\in I$. 

The paths in $\PD^{s_\Goid}(\Goid)$, which commence at an identity of $\Goid$, are labeled by $\PD^{s_\Goid}_{\{\Goid^{u}_u\}}(\Goid)$. 
Every element of $\tg\in\PD^{s_\Goid}(\Goid)$ is of the form $R_{\tg(0)}\cdot\tvt=\tg$, where $\tvt\in\PD_{\{\Goid^{u}_u\}}^{s_\Goid}(\Goid)$ for $u=\tg(0)$ and $R_{\tg(0)}=\ta$ for a loop $\ta$ in $\Goid^u_u$. This corresponds to a right-translation $R_{w}$ corresponding to $\Goid$ for $w\in\Goid^0$. 
In comparison to definition \ref{defi transfLie} one can also rewrite $R_{u}$ by $R_{\ta}$.

Consider a lift $\tg:\bra 0,1\ket\rightarrow \Goid^u$ such that $\tg\in\PD^{s_\Goid}_{\{\Goid^{u}_u\}}(\Goid)$ . Let $\ta\in\Goid^u_u$ be a lifted loop, then it is true that $\ta\circ\tg\in \PD^{s_\Goid}_\Sigma(\Goid)$, where
$\ta(0)\in\Goid^u_u$ such that $(\ta\circ\tg)(0)\in\Goid^u_u$, $\tg_t(1)\in\Goid^u_p$, $(\ta\circ\tg)(1)\in\Goid^u_p$. 
Moreover the surjections satisfy
$t_{\PD\Goid}(\tg(s))=\gamma(t)$, $t_\Goid(\idf_v)=\gamma(0)=u$  and $t_\Goid(\tg(1))=\gamma(1)=w$.

Now consider the following picture, which illustrate the definition of a path connection.
\begin{center}
\includegraphics[width=0.4\textwidth]{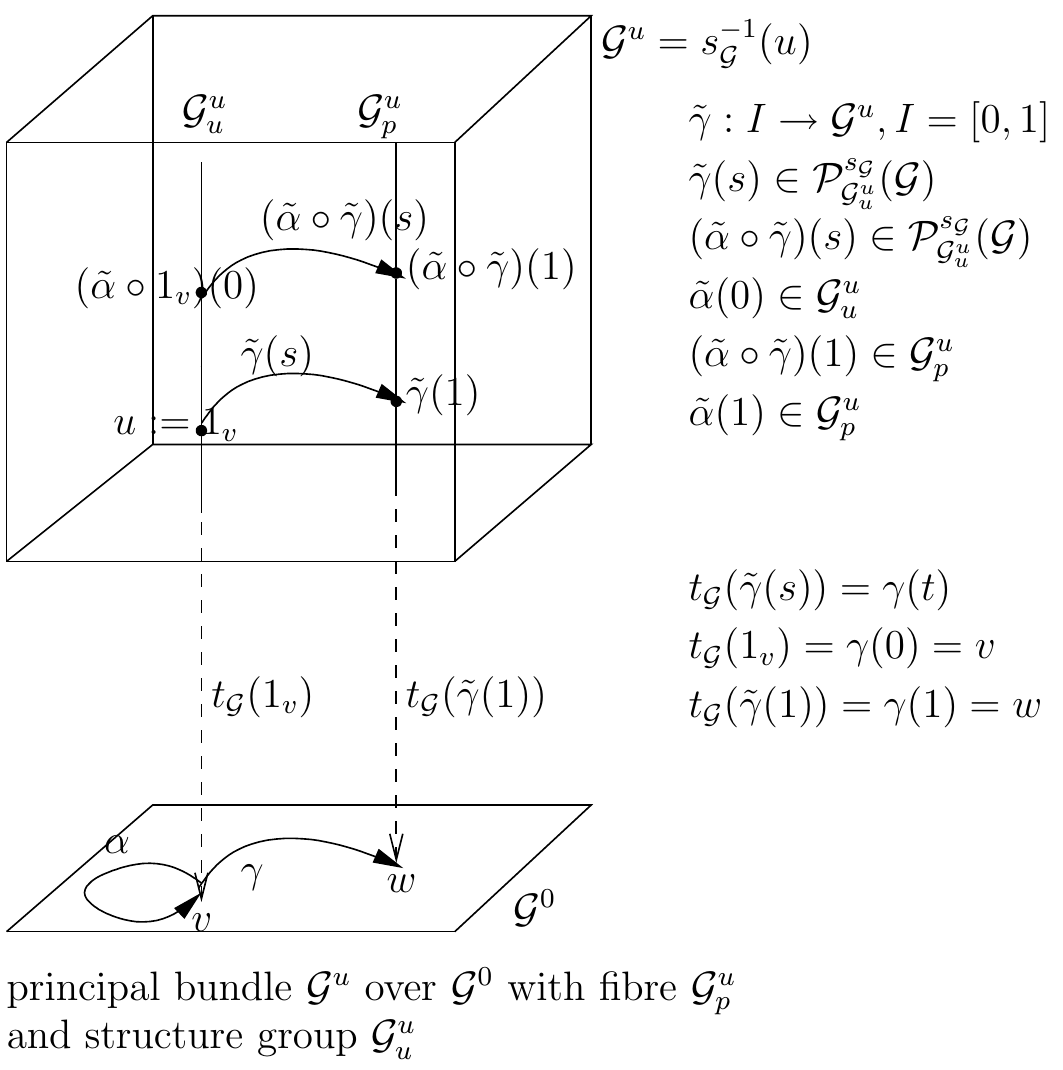}
\end{center}

\begin{defi}\label{def_pathconnectionLiegroupoid}
Let $\Goid$ be a locally trivial Lie groupoid over a connected base $\Sigma$. 

A \hypertarget{Path-connection}{\textbf{path connection}} in a Lie groupoid $\Goid$ on a base space $\Sigma$ is a map 
\[\Lambda: P\Sigma\rightarrow \PD^{s_\Goid}_{\{\Goid^{v}_v\}}(\Goid),\quad 
\gamma\mapsto \Lambda(\gamma)=:\tg\] 
where $I\ni s\mapsto \Lambda(\gamma,s)=:\tg(s)\in\Goid^v$ and
$I\ni t\mapsto \Lambda(\gamma)(t)=:\tg_t\in \PD^{s_\Goid}_{\{\Goid^{u}_u\}}(\Goid)$
such that the following conditions are satisfied:
\begin{enumerate} 
\item\label{path connection1} \hypertarget{path-conn}{\textbf{Start and target condition}}: The $\Lambda$-lift $\tg$ of a path $\gamma$ into the bundle $\{\Goid^v\}$ start at $\Goid_{v}^{v}$ and it is constant, i.o.w.
\beq s_{\PD\Goid^v}(\Lambda(\gamma))=\idf_{\gamma(0)}\in P\Sigma^v_v\eq 
where $s_{\Goid}(\Lambda(\gamma))=\gamma(0)=:v$.
The projection $t_{\PD\Goid^v}:\PD_{\{\Goid^v_v\}}^{s_\Goid}(\Goid^v)\rightarrow P\Sigma$ of a path $\Lambda(\gamma)$ starting at $\Goid^v_v$ in the Lie groupoid $\Goid^v$ onto the base space $\Sigma$ is $\gamma$, 
\beq t_{\PD\Goid^v}(\Lambda(\gamma))=\tg(1)=:\gamma(t)\in P\Sigma^v_{w}\eq
and $t_\Goid(\Lambda(\gamma))=\gamma(1)=:w$.
Hence, $s\mapsto\Lambda(\gamma,s)$ is a path in $\Goid^{\gamma(0)}_{\gamma(t)}$.
 \item\label{Reparametrization1} \hypertarget{reparametrisation-con}{\textbf{Reparametrization}}:  for every diffeomorphism $\phi:I\rightarrow [a,b]\subset I$ there is a right-translation $R_v:\PD^{s_\Goid}(\Goid)\rightarrow \PD_{\{\Goid^v_v\}}^{s_\Goid}(\Goid)$ for every $v\in\Sigma$ such that for every path $\gamma:[0,1]\rightarrow \Sigma$
\beq \Lambda(\gamma\circ\phi)
=R_{\Lambda^{-1}(\tg)(\phi(0))}\cdot (\Lambda(\gamma)\circ\phi)\eq
where $\Lambda^{-1}:\PD_{\{\Goid^v_v\}}^{s_\Goid}(\Goid)\rightarrow P\Sigma$ and $I\ni t\mapsto \Lambda^{-1}(\tg)(t)\in\Sigma$.
\item\label{Smoothness1} \hypertarget{smoothness-con}{\textbf{Smoothness}}: if $\gamma\in\PD\Sigma$ is smooth at $t=t_0\in I$, then $\tg_t$ is also smooth at $t=t_0$
\item\label{Tangency1} \hypertarget{tangency-con}{\textbf{Tangency}}: $\gamma,\gp\in\PD\Sigma$ if the tangent vectors coincide at some $t_0\in I$, 
\beqs \frac{\dif \gamma(t)}{\dif t}\Big\vert_{t=t_0}=\frac{\dif \gp(t)}{\dif t}\Big\vert_{t=t_0}\text{ then } \frac{\dif \tg_t(s)}{\dif t}\Big\vert_{t=t_0}=\frac{\dif \tgp_t(s)}{\dif t}\Big\vert_{t=t_0}\eqs for $\tg_t(s)=\Lambda(\gamma,s)(t)$ and $\tgp_t(s)=\Lambda(\gp,s)(t)$ and every $s\in I$
\item\label{Additivity1} \hypertarget{additivity-con}{\textbf{Additivity}}: $\gamma,\gp,\gpp\in\PD\Sigma$ if the tangent vectors satisfy at some $t_0$, 
\beqs \frac{\dif \gamma(t)}{\dif t}\Big\vert_{t=t_0}+\frac{\dif \gp(t)}{\dif t}\Big\vert_{t=t_0}
=\frac{\dif \gpp(t)}{\dif t}\Big\vert_{t=t_0}\text{ then } 
\frac{\dif \tg_t(s)}{\dif t}\Big\vert_{t=t_0}+\frac{\dif \tgp_t(s)}{\dif t}\Big\vert_{t=t_0}
=\frac{\dif \tgpp_t(s)}{\dif t}\Big\vert_{t=t_0}\eqs for $\tg_t(s):=\Lambda(\gamma,s)(t)$, $\tgp_t(s):=\Lambda(\gp,s)(t)$ and $\tgpp_t(s):=\Lambda(\gpp,s)(t)$ and every $s\in I$.
\end{enumerate}
\end{defi}

Now, the path connection satisfies some basic properties.
\begin{prop}\cite[Prop. 6.3.3]{Mack05} Let $\Lambda$ be a path connection in a Lie groupoid $\Goid$ over $\Goid^0$.

Then 
\begin{enumerate}
 \item\hypertarget{Unit-con}{Unit preserving}: $\Lambda(\tilde\idf_v)=\idf_{v}$ where $\tilde\idf_v$ is the constant path at $v$ in $\PD\Sigma$ and $\idf_v$ the constant path in $\PD\Goid$ where $\pi(\idf_v)=v$.
 \item\hypertarget{Inverse-con}{Inverse preserving}: $\Lambda(\gamma^{-1})=(R_{\Lambda^{-1}(\tg)(1)}\cdot\Lambda)(\gamma)^{\leftarrow}$ where $\gamma(t)^{-1}=\gamma(1-t)$, $\Lambda(\gamma)^{\leftarrow}(s)=\Lambda(\gamma,s-1)$ are the reversal paths and $R_v:\PD^{s_\Goid}(\Goid)\rightarrow \PD_{\{\Goid^v_v\}}^{s_\Goid}(\Goid)$
 \item\hypertarget{Concatenation-con}{Concatenation of paths}: $\Lambda(\gamma\circ\gp)=(R_{\tgp(1)}\cdot\Lambda(\gamma))\circ\Lambda(\gp)$ for every $s\in I$ and where $R_\gp:\PD_{\{\Goid^v_v\}}^{s_\Goid}(\Goid)\longrightarrow \PD^{s_\Goid}(\Goid)$ such that $s_{\PD\Goid^v}((R_{\tgp(1)}\cdot\tg)\circ \tgp)=\idf_{(\gamma\circ\gp)(0)}$ and $t_{\PD\Goid^v}((R_{\tgp(1)}\cdot\tg)\circ \tgp)=\gamma\circ\gp$.
\end{enumerate}
\end{prop}

These properties imply the following definition.
\begin{defi}For $\gamma\in P\Sigma$ the element $\ho_\Lambda(\gamma):=\Lambda(\gamma,1)\in\Goid^v_w$ where $v=\gamma(0)$ and $w=\gamma(1)$ is the \textbf{holonomy map of the path $\gamma$ in a Lie groupoid $\Goid$ over $\Goid^0$}.
\end{defi}

\begin{prop}\label{Path conn. groupoid} Let $\Lambda$ be a path connection in a groupoid $\Goid$ over $\Sigma$. 

Then
\begin{enumerate}
 \item\hypertarget{Unit-hol}{Unit}: $\ho_\Lambda(\idf_v)=e_\Goid(v)$  for $\idf_v$ the constant function in $P\Sigma$ and $e_\Goid(v)$ the constant function in $\Goid$ whenever $v\in\Sigma$,
 \item\hypertarget{Inverse-hol}{Inverse}: $\ho_\Lambda(\gamma^{-1})=\ho_\Lambda^{-1}(\gamma)$ for $\gamma\in P\Sigma$ and
 \item\hypertarget{Concatenation-hol}{Concatenation of paths}: $\ho_\Lambda(\gamma\circ\gp)=\ho_\Lambda(\gamma)\ho_\Lambda(\gp)$ if $t(\gamma)=s(\gp)$ and $\gamma,\gp \in P\Sigma$.
\end{enumerate}
\end{prop}
\begin{defi} The set $\Hol_\Lambda(\Sigma)=\{\ho_\Lambda(\gamma):\gamma\in P\Sigma\}$ is the \hypertarget{hol-groupoid}{\textbf{holonomy groupoid of $\Lambda$ associated to a groupoid $\Goid$ over $\Goid^0$}}. The vertex group $\Hol_\Lambda(\Sigma,v)$ at $v\in\Sigma$ is the \textbf{holonomy group of $\Lambda$ at $v$}. The vertex bundle $\{\Hol_\Lambda(\Sigma,v)\}$ is the \textbf{holonomy group bundle of $\Lambda$}
\end{defi}

For a Lie groupoid $\Goid$ over $\Goid^0$ the holonomy groupoid is a transitive subgroupoid of $\Goid$.

\begin{theo}\cite[Theorem 6.3.19]{Mack05} Let $\Lambda$ be a path connection in a \hyperlink{loc-triv-groupoid}{locally trivial} Lie groupoid $\Goid$.
 
Then $\Hol_\Lambda(\Sigma)$ is the \hypertarget{hol-groupoid}{holonomy Lie subgroupoid of $\Goid$}, the vertex group $\Hol_\Lambda(\Sigma,v)$ is the holonomy Lie group and the vertex bundle $\{\Hol_\Lambda(\Sigma,v)\}$ is the holonomy Lie group bundle of $\Lambda$. 
\end{theo}
Denote $H_\Lambda(\Sigma):=\nicefrac{\Hol_\Lambda(\Sigma)}{\ker \ho_\Lambda}$ and $H^{\breve\Lambda}_{\Lambda}(\Sigma):=\nicefrac{\Hol_\Lambda(\Sigma)}{\bigcap_{\Lambda\in\breve\Lambda}\ker \ho_\Lambda}$. 

\begin{defi}\label{def samehol}
Two paths $\gamma,\gp\in\PD\Sigma^v_w$ are said to have the \hypertarget{same-hol-conn}{\textbf{same-holonomy w.r.t. a fixed path connection $\Lambda$}} iff 
\beq\label{samehol} h_\Lambda(\gamma\circ\gp^{-1})=\idf_v \text{ where }s_{\PD\Sigma}(\gamma\circ\gp^{-1})=v, \pi(u)=v\eq 

Two paths $\gamma,\gp\in P\Sigma^v_w$ are said to have the \hypertarget{same-hol}{\textbf{same-holonomy w.r.t. all path connections}} iff \eqref{samehol} is true for all path connections $\Lambda\in\breve\Lambda$. 

Denote this relation by $\sim_{\text{s.hol. }\breve \Lambda}$.
\end{defi} 
This is obviously an equivalence relation. The consider the \textbf{hoop groupoid} $\Hoop_{\breve\Lambda}(\Sigma)=\{\ho_\Lambda(\gamma):\gamma\in\PD\Sigma /\sim_{\text{s.hol. }\breve \Lambda}\}$. If the groupoid $\Goid$ over $\Goid^0$ is equal to a connected Lie group $G$ over $\{e_G\}$, then the set $H_\Lambda^{\breve\Lambda}(\Sigma)$ and $\Hoop_{\breve\Lambda}(\Sigma)$ coincide. 

\begin{defi}
Let $\PD\Sigma$ be the path groupoid modulo same holonomy w.r.t. all path connections in $\breve \Lambda$. Moreover let $\Goid$ be a Lie groupoid over $\Goid^0$.

Then the groupoid morphism $\ho_\Lambda:\PD\Sigma\longrightarrow\Goid$ such that $\ho_\Lambda(\gamma)=\Lambda(\gamma,1)$ for all $\gamma\in\PD\Sigma$ associated to a path connection $\Lambda$ is called a \textbf{holonomy map in a Lie groupoid $\Goid$ over $\Goid^0$}. 

The \textbf{set of holonomy maps for a path groupoid} $\PGs$ in a Lie groupoid $\Goid$ over $\Goid^0$ is defined by 
\beqs\A_{\breve \Lambda,\Goid}:=\Hom_{\breve\Lambda}(\PD\Sigma,\Goid)=\{(\ho_\Lambda,h_\Lambda) \vert \qquad&\ho_\Lambda:\PD\Sigma\rightarrow \Goid, h_\Lambda:\Sigma\rightarrow\Goid^{(0)}\text{ such that }\\& (\ho_\Lambda,h_\Lambda)\text{ is a holonomy map for the path groupoid }\PGs\\
&\text{  and }\Lambda\in\breve\Lambda\}
\eqs
where $\breve\Lambda$ denotes the set of all path connections.  
\end{defi}

Notice that, the set $\breve \Lambda$ corresponds to the space of smooth connections $\breve\A_s$. Hence, for the particular Lie groupoid $G$ over $\{e_G\}$ the set $\A_{\breve \Lambda,G}$ is abrreviated by $\A_{s}$. 

Consider the example of the holonomy map constructed from the fundamental group $ \pi_1(\Sigma,v)$, which is a group homomorphism
\beqs \ho_A: \pi_1(\Sigma,v)\rightarrow G\eqs
Obviously, two paths $\gamma,\gp\in\Pi_1(\Sigma)^v_w$ have the same-holonomy w.r.t. a smooth connection $A$ in the set $\breve \A_s$ of smooth connections iff 
\beqs \ho_A(\gamma\circ\gp^{-1})=e_G\text{ and where }\Goid^v_v\simeq G\text{ and }e_G\text{ is the unit of the group } G\text{.}\eqs In other words, the loop $\gamma\circ\gp^{-1}$ is contractible to the constant loop at $v$. 

For $\Goid$ choose for example $\Pi_1(\Sigma)$ or $\Sigma\times\Sigma$. In fact, the fundamental groupoid $\Pi_1(\Sigma)$ is the biggest Lie groupoid and the pair groupoid the smallest among all Lie groupoid such that the corresponding Lie algebroid $\A\Goid$ is equivalent to $T\Sigma$. However, in the case of a gauge theory it is more interesting to consider the gauge groupoid $\fG$. Before the gauge groupoid is analysed in detail, some further properties are collected.

\paragraph*{Duality and the generalised Ambrose-Singer theorem\\[10pt]}

In the context of the gauge groupoid, the duality of infinitesimal objects and path connections and holonomies is based on the following theorem. 
\begin{theo}\label{theo ambrose}\cite[theorem 6.3.5]{Mack05})\\
There is a bijective correspondence between a path connection $\Lambda$ in the gauge groupoid $\fG$ and an infinitesimal Lie algebroid connection $\gamma_A:T\Sigma\rightarrow \frac{TP}{G}$ such that
\beqs \frac{\dif}{\dif t}\Big\vert_{t=t_0}\tg_t= T(R_{\Lambda(\gamma)(t_0)})\left(\gamma_A\Big(\frac{\dif}{\dif t}\Big\vert_{t=t_0}\gamma(t)\Big)\right)\eqs
whenever $\gamma\in P\Sigma$ and $\tg=\Lambda(\gamma)$.
\end{theo}

For $X\in T_v\Sigma$ and a path $\gamma\in P\Sigma$ with $\gamma(t_0)=v$ and $\frac{\dif}{\dif t}\Big\vert_{t=t_0}\gamma(t)=X$ for some $t_0\in I$ define
\beq \gamma_A(X):=T (R_{\Lambda^{-1}(\tg)(t_0)}) \Big(\frac{\dif}{\dif t}\Big\vert_{t=t_0}\tg_t\Big)
\eq

Note that, the theorem has been originally stated by Mackenzie \cite{Mack05} in the general context of transitive Lie groupoids and Lie algebroids. In this article this theorem is formulated in the context of gauge theories for a comparison with Barrett \cite{Barrett91}. The generalisation is deduced by replacing the Lie algeboid $\frac{TP }{G}$ by the transitive Lie algebroid $A\Goid$ associated to a transitive Lie groupoid $\Goid$.

There exists a \textbf{generalised exponential map} $\Exp:\Gamma A\Goid\longrightarrow \Gamma\Goid$ such that $\tilde X\mapsto \Exp(t\tilde X(v))$ for all $t\in\R$ and $v\in\Goid^0$. A local one-parameter group of (local) diffeomorphisms is given by the diffeomorphic maps $\phi_t:U\times \bra -\epsilon,\epsilon\ket\longrightarrow \Sigma$. Then a \textbf{local one-parameter group of (local) diffeomorphisms  $\tilde\varphi_t$ on the Lie groupoid $\fG$} with respect to $\phi_t$ is given by the gauge and diffeomorphism transformation $\tilde\varphi_t(\phi_t,\id_G)$ in $\fG$. Note that $\id_G$ is the identity map on $G$. Then the map $\R\ni t\mapsto t_P(\exp(t\tilde X(v)))$ defines a local one-parameter group, too. The map $(t,v)\mapsto \Exp(t\tilde X(v))$ is a \textbf{family of local bisections}.
Note that, it is true that $\tilde X(v)=(\gamma_A(X))(v)$ for $\gamma_A:T\Sigma\longrightarrow A\Goid$, where $A\Goid:=\frac{TP}{G}$, holds.
\begin{cor}
 Let $\gamma_A$ be an infinitesimal connection in $\Goid$ and let $\Lambda$ be the corresponding path connection. Let $\{\varphi_t(v)\}$ be a local flow for a vector field $X$ near $v$. Set $\varphi(t):=\varphi_t(v)$.

Then 
\beqs\Exp(t(\gamma_A(X))(v))=\Lambda(\varphi,v)(t)
\eqs where $\Lambda(\varphi,v):\R\longrightarrow \Goid^v$ is the lift of the path $t\mapsto \varphi_t(v)$.
\end{cor}

Furthermore, the next theorem give a correspondence between an object constructed from curvature forms, infinitesimal connections and the holonomy groupoids associated to path connections in the context of a gauge theory.

\begin{theo}\label{Theo GenAmbroseSinger}(\textbf{Generalised Ambrose-Singer theorem} \cite[theorem 6.4.20]{Mack05})\\
Let $P(\Sigma,G)$ be a principal bundle and $A$ an infinitesimal connection in the gauge groupoid $\fG$ associated to a path connection $\Lambda$. 

Then there exists a least Lie subalgebroid of the Lie algebroid constructed from $P(\Sigma,G)$ which contains the values of the right splitting $\gamma_A$ associated to $A$ and the values of its curvature form, and this Lie subalgebroid is the Lie algebroid corresponding to the holonomy groupoid $\Hol_\Lambda(\Sigma)$ of $\Lambda$. 
\end{theo}

Explicitly, in the example of the gauge groupoid $\fG$ this theorem has the following form. Let $\ko$ be the Lie subalgebra of $\go$ generated by $\{\Omega(X,Y)\in\Omega^2(P,\go): X,Y\in TP\}$. Then $\ko$ is the least Lie subalgebroid of the Lie algebroid constructed from $P(\Sigma,G)$. The Lie algebroid $\{\Omega(X,Y)\in\Omega^2(P,\go): X,Y\in TP\}$ can be related to a Lie algebroid subbundle of $\frac{P\times\go}{G}$. 

The theorem \ref{Theo GenAmbroseSinger} has been originally stated by Mackenzie \cite{Mack05} in the general context of transitive Lie groupoids and Lie algebroids. The generalisation can be obtained by replacing the Lie algeboid $\frac{TP }{G}$ by the transitive Lie algebroid $A\Goid$ associated to a transitive Lie groupoid $\Goid$  over $\Sigma$.

\subsection{The quantum configuration variables for a gauge theory}\label{sec Groupmorphism}
The simpliest holonomy maps are group homomorphisms from the (thin/intimate) fundamental group to a chosen Lie group. The concept is enlarged by holonomy maps, which are groupoid morphisms. Therefore, the concepts of holonomy maps, which have been presented by Barrett in \cite{Barrett91}, has been rewritten in the last section such that in comparison with the work of Mackenzie the holonomy maps for different physical theories are generalised. In \cite[Section 3.3.1 and 3.3.2]{KaminskiPHD} it has been shown that, the theory of Mackenzie really generalises the examples of Barrett for Yang Mills and gravitational theories. In this article certain holonomy maps for the example of a general gauge theory are presented in section \ref{subsec holmapsgaugetheories2}. This formulation extends the work of Barrett for Yang Mills theories and is an application of Mackenzie's theory for Lie groupoids. 

In \cite{Kaminski1,Kaminski2,Kaminski3,Kaminski4}, \cite[Sec.: 3.3.4.1, 3.3.4.3]{KaminskiPHD} the ideas has been generalised in the context of (semi-)analytic paths. The concept of holonomy maps (in the sense of Barrett) for finite path groupoids has been reformulated and has been further generalised to the case of finite graph  systems. The generalisation of Barrett's objects to the framework of Mackenzie's path connections forces to generalise the holonomy mappings for a finite path groupoid once more. Indeed the set of conditions that a holonomy map is required to satisfy has been extended in \cite[Sec.: 3.3.4.2]{KaminskiPHD}.

\subsubsection{Holonomy maps and transformations for a gauge theory}\label{subsec holmapsgaugetheories2}

In this section the concept of holonomy maps of Barrett is further generalised with the help of the ideas of Mackenzie. In particular, the holonomy mappings map paths to elements of the gauge groupoid instead of elements of the structure group of a principal fibre bundle. This lead to a new formulation of the configuration space of Loop Quantum Gravity: the space of smooth conncetions correspond one-to-one to the space of holonomy maps for a gauge theory.

In the first subsection the definition of holonomy maps is considered for the certain case of a general gauge theory. Then gauge and diffeomorphism transformations on the holonomy groupoid associated to a path connection are studied explicity. Finally the new formulation of the configuration and momentum space is presented.

\paragraph*{The holonomy maps and the holonomy groupoid for a gauge theory\\[10pt]}

Consider a principal bundle $P(\Sigma,G)$ and construct the gauge groupoid $\fG$.
Moreover recall
\beqs\Goid^v&=\{\langle u,p\rangle: \pi(u)=v\}\\
\Goid^v_v&=\{\langle u,p\rangle: \pi(u)=v=\pi(p)\}\\
&=\{\langle u\delta(p,u),u\rangle\}
\eqs

Then a path connection in the Lie groupoid $\fG$ is a map $\Lambda_u: \PD\Sigma^v\rightarrow \PD^{s_\Goid}_{\{\Goid^{v}_v\}}\Goid^v$ for fixed $u\in P$, where $\pi(u)=v$, which is given by
\beqs \Lambda_u: \gamma \mapsto \Lambda_{u}(\gamma):=\langle u ,\tg_t\rangle\text{ for } \gamma\in \PD\Sigma^v
\eqs
where the map $s\mapsto \tg(ts)=:\tg_t(s)$ is a lifted path in $P$ of a path $t\mapsto\gamma(t)$ in $\Sigma$ such that $\tg(0)=u$ and $\tg(t)=p(t)$, $\pi(p(t))=\gamma(t)$. Rewrite $\Lambda(\gamma)(s)=:\Lambda(\gamma,s)$ and $\Lambda(\gamma,s):=\la u,\tg(ts)\ra$. Hence for a fixed path $\gamma$ the mapping $s\mapsto \Lambda(\gamma,s)$ maps from an intervall $I$ to $\Goid^v$.
The source and target maps satisfy
\beqs s_P( \langle u,\tg(ts)\rangle)&:=(\pi\circ\pr_1)( \langle u ,\tg(ts)\rangle)\\&= \pi(u)=v\eqs
\beqs t_P(\langle u ,\tg(ts)\rangle)&:=(\pi\circ\pr_2)(\langle u ,\tg(ts)\rangle)\\&= \pi(\tg(ts))= \gamma(ts)=(\gamma\circ\varrho_s)(t) \eqs
where
\beq\langle u, u\rangle =\idf_v\text{ and } t_P(\langle u, u\rangle)=t_P(\idf_v)=\pi(u)= v=s_P(\idf_v)\eq
Summarising the path-connection $\Lambda$ depends on the choice of the path $\gamma$ and the  point $u\in P$. 

The reversal $\Lambda(\gamma,s)^\leftarrow$ is given by
\beqs \Lambda(\gamma,s)^\leftarrow:=\la \tg(t),u\ra\la u,\tg(t(1-s))\ra=\la \tg(t),\tg(t(1-s))\ra\eqs

The concatenation operation $\cdot$ is defined by
\beq &(\Lambda^1_{u}\cdot\Lambda^2_{p})(s) 
:= R_{\gamma_2}\circ\langle u,\tg(t)(s)\rangle\cdot\Lambda^2_{p}(\gamma_2)
=\langle u,(\tg_1\circ\tg_2)(ts)\rangle \la p,\tg_2(ts)\ra \\
&=\left\{\begin{array}{cc}
\la u,\tg_1(2ts)\ra\la p,\tg_2(ts)\ra & \text{ for } s\in \bra 0,\nicefrac{1}{2}\ket\\
\langle u,\tg_2(t(2s-1))\ra\la p,\tg_2(ts)\ra & \text{ for } s\in \bra \nicefrac{1}{2},1\ket\\
\la u,\tg_1(0)\ra\la p,\tg_2(0)\ra = \la u,u\ra\la p,p\ra&\text{ for } s=0\\
\langle u,\tg_2(t)\ra=\langle \tg_1(0),\tg_2(t)\ra  &\text{ for } s=1\\
  \end{array}\right.
\eq
In general, the \textbf{holonomy map for a gauge theory} is a groupoid morphism presented by the map
\beq \Ho_\Lambda:\PD\Sigma^v\rightarrow \Goid^v, \gamma\mapsto \Ho_\Lambda(\gamma):=\Lambda\vert_{s=1}(\gamma)=\la u,\tg_t(1)\ra\eq

The \textbf{holonomy groupoid for a gauge theory} is given by
\beqs \Hol_\Lambda^P(\Sigma):=\{\Ho_\Lambda(\gamma): \gamma\in\PD\Sigma\}
\eqs

Moreover, for a loop $\alpha\in\PD\Sigma^v_v$ it is true that
\beq\Lambda(\alpha,s)\vert_{s=1}=\langle u ,\ta_t(1)\rangle=\langle u ,\ta(t)\rangle
\eq where $\ta_t(1):=(\ta\circ\varrho_s)\vert_{s=1}(t)=\ta(t)$.
In fact, there is a unique element $\ho_\Lambda(\alpha)\in G$ such that
\beq \langle u ,\ta(t)\rangle\langle u,u\rangle =\langle u\delta(\ta(t),u),u\rangle\text{ for }\pi(\ta(t))=\pi(u)=v\eq where $\delta(\ta(t),u) =: \ho_\Lambda(\alpha)$.
Consequently, the holonomy maps are related to the holonomy maps $\ho_\Lambda: \PD\Sigma^v_v\rightarrow G$. 

The lift $\tg(t)$ of a path $\gamma(t)$ in $P$ is a diffeomorphism between the fibres $P_v$ and $P_{\gamma(t)}$, i.e. the map $\tg(t): I \rightarrow \Diff(P_v,P_{\gamma(t)})$. 

A change of the base point $u\in P$ transforms as follows
\beq \Lambda_{uk}(\gamma,s)&=\langle uk ,\tg_k(ts)\rangle\\
&=\langle uk,\tg_k(ts)k^{-1}k\rangle=\langle uk,\tg(ts)k\rangle =R_k\circ \Lambda_u(s)\\
&=\la u, \tg(ts) \ra = \Lambda_u(\gamma,s)
\eq where $\tg_k(0)=uk$ and $\tg(0)=u$ such that
\beq \Lambda_{uk}(\gamma,1)\idf_v=\langle uk,\tg(t)k\rangle\la u,u\ra= \la u \Ad(k)(\ho_\Lambda(\gamma)),u\ra= \la u \ho_\Lambda(\gamma),u\ra =\Lambda_u(\gamma,1)\idf_v
\eq
This means that the lifts $\tg$ have to be $G$-compatible, w.o.w. $\tg_k(ts)k^{-1}=\tg_{kk^{-1}}(ts)=\tg(ts)$ for all $k\in G$. Summarising the lifts $\tg$ are contained in the set $\Diff_{G\text{-comp.}}(P_v,P_w)$ of diffeomorphism between the fibres $P_v$ and $P_{\gamma(t)}$, which are $G$-compatible.

\paragraph*{Transformations in holonomy groupoid for a gauge theory\\[10pt]}

Let $\Goid:=\frac{P\times P}{G}$  over $\Sigma$ be the gauge groupoid. Then the holonomy groupoid $\Hol_\Lambda^P(\Sigma)$ has been derived in the last section from a holonomy map $\Ho_\Lambda:P\Sigma\rightarrow \Goid$, which has been defined by 
\beqs \Ho_\Lambda(\gamma)=\langle u ,\tg_t(1)\rangle= \la u,\tg(t)\ra
\eqs where $\tg_t(1):=(\tg\circ\varrho_s)\vert_{s=1}(t)=\tg(t)$ for a lifted path $s\mapsto \tg(ts)$ in $P$ of a path $\gamma\in P\Sigma$ and $s_{P\Sigma}(\gamma)=v$, $\pi(v)=u=s_{P}(\tg)$.

\begin{defi}
Let $\sigma(v):=\la u,\varphi(u)\ra$ defines a bisection $\sigma$ of $\fG$ for a gauge and diffeomorphism transformation $\varphi(\varphi_0,\id)$, where $\pi(\varphi(u))=\varphi_0(\pi(u))=\varphi_0(v)$ holds. 

Then a left action $L_\sigma$ on the holonomy groupoid $\Hol_\Lambda^P(\Sigma)$ is defined by
\beqs L_\sigma\Ho_\Lambda(\gamma):=\sigma((t_P\circ\sigma)^{-1}(s_P(\Ho_\Lambda(\gamma)))) \Ho_\Lambda(\gamma)= \la u,\varphi(u)\ra \Ho_\Lambda(\gamma)
\eqs whenever $\Ho_\Lambda(\gamma)\in\Hol^P_\Lambda(\Sigma)$.
\end{defi}
This holds since $\sigma(\pi(u))=\sigma(v)=\la u,\varphi(u)\ra$ and $(t_P\circ\sigma)(v)=\pi(\varphi(u))=\varphi_0(v)$. Notice that, $(\pi\circ s_P)(L_\sigma\Ho_\Lambda(\gamma))=v$ and $t_P(L_\sigma\Ho_\Lambda(\gamma))=\tg(1)$ yields.

For pure gauge transformations, w.o.w. if $\pi(\varphi(u))=\pi(u)$ and $\varphi_0(v)=v$ is satisfied, derive
\beq L_\sigma\Ho_\Lambda(\gamma)=\la u,\varphi(u)\ra\la u,\tg_t(1)\ra =\la u\delta(\varphi(u),u),\tg_t(1)\ra 
\eq In particular pure gauge transformations are given by $\varphi(u):=ug$ for suitable $g$ in $G$. Then since $\varphi$ is required to satisfy $\varphi(uk)=\varphi(u)k$ for all $k\in G$, the element $g\in G$ have to be such that $[g,k]=0$ for all $k\in G$. Denote the center of the group $G$ by $\ZD(G)$. Hence for pure gauge transformations there is an action $L$ on $\fG$ defined by
\beqs 
&L:\Big(\frac{P\times P}{G}, (P\times \ZD(G))\Big)\longrightarrow \frac{P\times P}{G};\\
&L_{(u_2,g)}(\la p,u_1\ra):= 
\left\{\begin{array}{ll}
\la p,u_1\ra \ast (u_2,g)=\la p,u_1g\ra  & \text{ for }\pi(u_1)=\pi(u_2)=v \\
\la p,u_1\ra \ast (u_2,g)=\la p,u_1\ra  & \text{ for }\pi(u_1)\neq\pi(u_2) \\
\end{array}\right.
\eqs  Therefore the left action on $\Hol^P_\Lambda(\Sigma)$ associated to a bisection $\sigma$, which is build from a pure gauge transformation, is presented by
\beq &L_\sigma\Ho_\Lambda(\gamma)=\la u,\varphi(u)\ra\la u,\tg_t(1)\ra= (\la u,u\ra\ast (u,g))\la u,\tg_t(1)\ra 
\\&=\la u,ug\ra\la u,\tg_t(1)\ra =\la u\delta(ug,u),\tg_t(1)\ra =\la u g^{-1},\tg_t(1)\ra \eq and it is true that
\beq L_\sigma\Ho_\Lambda(\gamma)&=\la uk,ukg\ra\la u,\tg_t(1)\ra= \la uk,ugk\ra\la u,\tg_t(1)\ra =\la uk\delta(ugk,u),\tg_t(1)\ra \\&=\la u g^{-1},\tg_t(1)\ra \text{ for all } k\in G\eq  holds.

Since $\tg_t(1)g=:\tg_g(1)$ and $\tg(0)g=ug$ holds, deduce 
\beq L_\sigma\Ho_\Lambda(\gamma)=\la u,ug\ra\la u,\tg_t(1)\ra &=\la u ,\tg_t(1)g\ra=\la u ,\tg_g(1)\ra \\
\eq

\begin{defi}Define the action $L_{\sigma^{-1}}$ on $\Hol_\Lambda^P(\Sigma)$ for a bisection $\sigma^{-1}$ of $\Hol_\Lambda^P(\Sigma)$ by 
\beqs L_{\sigma^{-1}}\Ho_\Lambda(\gamma)&:= \sigma^{-1}(s_P(\Ho_\Lambda(\gamma))) \Ho_\Lambda(\gamma)= \la \varphi(u),u\ra \Ho_\Lambda(\gamma)\\
&= \la \varphi(u),\tg_t(1)\ra
\eqs for every gauge and diffeomorphism transformation $\varphi(\varphi_0,\id)$ and $\pi(u)=v$. 
\end{defi}
Notice that, $(\pi\circ s_P)(\Ho_\Lambda(\gamma))=v$, $(\pi\circ s_P)(L_{\sigma^{-1}}\Ho_\Lambda(\gamma))=w$ holds, where $w:=(\pi\circ\varphi)(u)=\varphi_0(v)$ is not necessarily equal to $v$ and $t_P(L_{\sigma^{-1}}\Ho_\Lambda(\gamma))=\tg_t(1)$ is satisfied.

If $\pi(\tg_t(1))=\pi(u)=v$ is fulfilled, then 
\beq L_{\sigma^{-1}}\Ho_\Lambda(\gamma)\idf_v&= \la \varphi(u),\tg_t(1)\ra\la u,u\ra =\la \varphi(u)\delta(\tg_t(1),u),u\ra
\eq 
and $\ho_\Lambda(\gamma)=\delta(\tg_t(1),u)$ yields. If $\varphi(u)=u k$ for $k\in \ZD(G)$ is satisfied, then it follows that
\beq L_{\sigma^{-1}}\Ho_\Lambda(\gamma)\idf_v&= \la u k \ho_\Lambda(\gamma),u\ra
\eq holds.

In general the composition of paths transfers to multiplication of elements on $\frac{P\times P}{G}$:
\beq &\Ho_\Lambda(\gamma_1)L_{\sigma^{-1}}\Ho_\Lambda(\gamma_2)= \la u,\tg^1_t(1)\ra\la \varphi(u),\tg^2_t(1)\ra =\la u\delta(\tg^1_t(1),\varphi(u)),\tg^2_t(1)\ra
\eq Moreover if $\gamma^1(1)=\gamma^2(0)$ and $\pi(\gamma^2(0))=\pi(u)=v=\pi(\gamma^1(1))=\pi(\gamma^1(0))$ is true, then compute
\beq \Ho_\Lambda(\gamma_1)\Ho_\Lambda(\gamma_2)&= \la u,\tg^1_t(1)\ra\la u,\tg^2_t(1)\ra =\la u\delta(\tg^1_t(1),u),\tg^2_t(1)\ra\\&=\la u\ho_\Lambda(\gamma_1),\tg^2_t(1)\ra
\eq For a pure gauge transformation $\varphi(u)=ug$ for $g\in \ZD(G)$ then derive
\beq &\Ho_\Lambda(\gamma_1)L_{\sigma^{-1}}\Ho_\Lambda(\gamma_2)= \la u\delta(\tg^1_t(1),ug),\tg^2_t(1)\ra =\la u g\ho_\Lambda(\gamma_1),\tg^2_t(1)\ra
\eq If $\pi(\tg^2_t(1))=\pi(u)=v$ is fulfilled, then finally calculate
\beq &\Ho_\Lambda(\gamma_1)L_{\sigma^{-1}}\Ho_\Lambda(\gamma_2)\idf_v= \la u g\ho_\Lambda(\gamma_1)\ho_\Lambda(\gamma_2),u\ra
\eq Notice that, 
\beq &L_{\sigma^{-1}}\Ho_\Lambda(\gamma_1)\Ho_\Lambda(\gamma_2)\idf_v= \la u \ho_\Lambda(\gamma_1)g^{-1}\ho_\Lambda(\gamma_2),u\ra
\eq holds.

Recall that $\langle \tg_t(1)g,\varphi(\tg_t(1))g\ra=\la \tg_t(1),\varphi(\tg_t(1))$ and $(\varphi_0\circ\pi)(\tg_t(1))=(\pi\circ\varphi)(\tg_t(1))$ holds.
\begin{defi}
Let $\sigma(v):=\la u,\varphi(u)\ra$ defines a bisection $\sigma$ of $\fG$ for a gauge and diffeomorphism transformation $\varphi(\varphi_0,\id)$. 

Then a right action $R_\sigma$ on the holonomy groupoid $\Hol_\Lambda^P(\Sigma)$ is defined by
\beqs &R_\sigma\Ho_\Lambda(\gamma):= \Ho_\Lambda(\gamma)\sigma(t_P(\Ho_\Lambda(\gamma)))
=\la u,\tg_t(1)\ra\la \tg_t(1),\varphi(\tg_t(1))\ra \\
&=\la u,\varphi(\tg_t(1))\ra
\eqs
\end{defi}Then $(s_P\circ\pi)(R_\sigma\Ho_\Lambda(\gamma))=u=\pi(v)$ and $t_P(R_\sigma\Ho_\Lambda(\gamma))=\varphi(\tg_t(1))$ holds.
Observe that 
\beq &R_\sigma\Ho_\Lambda(\gamma) \idf_v=\la u,\varphi(\tg_t(1))\ra\la u,u\ra= \la u\delta(\varphi(\tg_t(1)),u),u\ra
\eq is true.
If pure gauge transformations are considered, i.e. $\varphi(\tg_t(1))=\tg_t(1) g$ for $g\in \ZD(G)$ holds, then derive
\beq &R_\sigma\Ho_\Lambda(\gamma) \idf_v=\la u,\tg_t(1)g\ra\la u,u\ra= \la u\delta(\tg_t(1)g,u),u\ra =\la u \ho_\Lambda(\gamma)g^{-1},u\ra
\eq Moreover for each $k\in G$ observe
\beq &R_\sigma\Ho_\Lambda(\gamma) \idf_v=\la uk,\tg_t(1)gk\ra\la u,u\ra= \la u \Ad(k)(\ho_\Lambda(\gamma))g^{-1},u\ra
\eq whenever $g\in \ZD(G)$.

If $\pi(\gamma_1)=\pi(u)=v$ is satisfied, then calculate
\beq &\Ho_\Lambda(\gamma_1)R_\sigma\Ho_\Lambda(\gamma_2)=\la u,\tg^1_t(1)\ra\la u,\varphi(\tg^2_t(1))\ra= \la u\delta(\tg^1_t(1),u),\varphi(\tg^2_t(1))\ra\\
&=\la u \ho_\Lambda(\gamma_1),\varphi(\tg^2_t(1))\ra
\eq Moreover if $\pi(\gamma_2)=\pi(u)=v$ and $\varphi(\tg^2_t(1))=\tg^2_t(1)g$ yields for $g\in \ZD(G)$, then derive
\beq &\Ho_\Lambda(\gamma_1)R_\sigma\Ho_\Lambda(\gamma_2)\idf_v= 
&=\la u \ho_\Lambda(\gamma_1)\delta(\varphi(\tg^2_t(1)),u),u\ra =\la u \ho_\Lambda(\gamma_1)\ho_\Lambda(\gamma_2)g^{-1},u\ra 
\eq

\begin{defi}
Let $\sigma(v):=\la u,\varphi(u)\ra$ defines a bisection $\sigma$ of $\fG$ for a gauge and diffeomorphism transformation $\varphi(\varphi_0,\id)$, where $\pi(\varphi(u))=\varphi_0(\pi(u))=\varphi_0(v)$ holds. 

Then an inner action $I_\sigma$ on the holonomy groupoid $\Hol_\Lambda^P(\Sigma)$ is implemented by 
\beqs I_\sigma(\Ho_\Lambda(\gamma)))&:= R_\sigma(L_{\sigma^{-1}}\Ho_\Lambda(\gamma))=  \sigma^{-1}(s_P(\Ho_\Lambda(\gamma)))\Ho_\Lambda(\gamma)\sigma(t_P(\Ho_\Lambda(\gamma)))\\
&=\la \varphi(u),u\ra \la u,\tg_t(1)\ra\la \tg_t(1),\varphi(\tg_t(1))\ra \\
&=\la \varphi(u),\tg_t(1)\ra\la \tg_t(1),\varphi(\tg_t(1))\ra= \la \varphi(u),\varphi(\tg_t(1))\ra 
\eqs whenever $\Ho_\Lambda(\gamma)\in\Hol^P_\Lambda(\Sigma)$.
\end{defi}
If additionally $\varphi(u)=ug$ for $g\in\ZD(G)$ holds, then compute
\beq &R_\sigma(L_{\sigma^{-1}}\Ho_\Lambda(\gamma))= \la u,\varphi(\tg_t(1))\ra 
\eq

Let $\la u,\tg\ra\in \PD\Goid^v$ a path in the gauge groupoid $\fG$, where $\pi(u)=v$ and $\tg\in\PD P$. Let $\rho_\tg:I\rightarrow G$ be a curve such that $\rho_\tg(0)=e_G$, $\rho_\tg(1)=g$ for $g\in G$ and $R_k(\rho_\tg(s))=\Ad(k)(\rho_\tg(s))$ for all $k\in G$. Then for a fixed $s\in I$ concern the following action associated to a purely gauge transformation as a map
\beqs&\ast: \Big(\frac{P\times P}{G},\frac{P\times G}{G}\Big)\longrightarrow \frac{P\times P}{G},\text{ where }\\
&\la u,\tg_1(s)\ra\ast \lfloor \tg_2(s),\rho_\tg(s)\rfloor
=\left\{\begin{array}{ll}
         \la u,\tg_t(s)\rho_\tg(s)\ra &\text{ for }(\pi\circ t_P)(\tg_1)=(\pi\circ t_P)(\tg_2)=v\\
	 \la u,\tg_1(s)\ra&\text{ for }(\pi\circ t_P)(\tg_1)\neq(\pi\circ t_P)(\tg_2)\\
        \end{array}\right.
\eqs Recall 
\[\lfloor \tg_t(s),\rho_\tg(s)\rfloor =\lfloor \tg_t(s)kk^{-1},\rho_\tg(s)\rfloor=\lfloor \tg_t(s)k,\Ad(k^{-1})(\rho_\tg(s))\rfloor\] 

Then the action $R$ of a bisection $\sigma$ associated to a purely gauge transformation on the gauge groupoid $\fG$ is reformulated by 
\beq R_\sigma\la u,\tg_t(s)\ra&= \la u,\varphi(\tg_t(s))\ra  
=\la u,\tg_t(s)\ra\ast \lfloor \tg_t(s),\rho_\tg(s)\rfloor
= \la u,\tg_t(s)\rho_\tg(s) \ra
\eq whenever $\tg_t(s)\in P$ for a fixed $s\in I$, $(\tg_t(s),\rho_\tg(s))\in \frac{P\times G}{G}$ and such that
\beq R_\sigma\la uk,\tg_t(s)k\ra&= \la uk,\varphi(\tg_t(s))k\ra= \la uk,\varphi(\tg_t(s)k)\ra = \la uk,\tg_t(s)k\Ad(k^{-1})(\rho_\tg(s))\ra\\
&=\la u,\tg_t(s)\rho_\tg(s)\ra
\eq yields.
Futhermore for an element $\la u,\tg_t(1)\ra\in\Hol_\Lambda^P(\Sigma)$ it is true that,
\beq R_\sigma\la uk,\tg_t(1)k\ra\idf_v&=\la uk,\tg_t(1)\rho_\tg(1)k\ra\la u,u\ra=\la uk \delta(\tg_t(1)\rho_\tg(1)k,u)u\ra \\&= \la u \Ad(k)(\ho_\Lambda(\gamma)) \rho_\tg(1)^{-1} ,u\ra
\eq holds
such that
\beq R_\sigma\la uk,\tg(0)k\ra=  \la u, u\ra\quad\forall k\in G
\eq is true. Choose the map $\rho_\tg$ such that
\beq\rho_\tg(s):=\exp(X_{\tg_t(s)})\text{ for }X_{\tg_t(s)}\in\go \text{ and }
\Ad(g)\rho_\tg(s)=\rho_\tg(s)
\eq for a fixed element $s\in I$ and where $\tg_t(s)\in P$ is satisfied.

Let $W$ be an open subset of $\Sigma$. Let $X_{\tg_t(1)}$ be an element of the section $\Gamma(\frac{T_WP}{G})$, then there exists a family of local bisections $\exp(tX_{\tg_t(1)})$ such that 
\beq\frac{\dif}{\dif \tau}\exp(\tau X_{\tg_t(1)})\Big\vert_{\tau=0} = X_{\tg_t(1)}\eq 
holds. Therefore set
\beq\rho_\tg(1):=\exp(\tau X_{\tg_t(1)})\text{ for }X_{\tg_t(1)}\in\go, \tau\in \R \eq
and require $\Ad(g^{-1})\rho_\tg(1)=\rho_\tg(1)$, where $\tg_t(1)\in P$ and $\pi(\tg_t(1))=w\in\Sigma$. 

Consequently there is a right action $R_{\sigma}$ on $\Hol_\Lambda(\Gamma)$ associated to the bisection $\sigma=(\varphi,\id_\Sigma)$, where $\varphi(\tg_t(s))= \tg_t(s) \rho_{\tg}(s)$ for all paths $\tg_t(s)$ in $P$.
Observe that, a connection reform $\omega^{/G}:\frac{TP}{G}\rightarrow\frac{P\times\go}{G}$ is a map, which satisfies\\
\[\omega^{/G}_{pg}(T(R_g)_pX_{\tg_t(1)})=\Ad(g^{-1})(\omega^{/G}_{p}(X_{\tg_t(1)}))=\omega^{/G}_{p}(X_{\tg_t(1)})\]
for $X_{\tg_t(1)}\in T_pP$ and $p=\tg_t(1)$.

\subsubsection{A set of holonomy maps for a gauge theory}\label{subsubsec holmapdef}

A smooth connection correspond one-to-one to a holonomy groupoid $\Hol^P_\Lambda(\Sigma)$. Assume that, the holonomy groupoid $\Hol^P_\Lambda(\Sigma)$ is a transitive Lie subgroupoid of the gauge groupoid $\fG$.
Recall the vector bundle $T_{e_G}(G)$ for a Lie group $G$. The elements of $T_{e_G}(G)$ are the right invariant vector fields. Therefore these right invariant vector fields $X$ correspond to generalised fluxes $E$, which are derivations on $G$.

The Lie algebroid associated to the holonomy Lie groupoid $\Hol_\Lambda^P(\Sigma)$ is given by 
\beqs A(\Hol_\Lambda^P(\Sigma)):= \left\{ X\in\frac{TP}{G}: X-(\gamma_A\circ\pi_*)(X)\in\Ima(\bar R_A) \right\}
\eqs 

The space $\breve\A_s$ of smooth connections is an affine space w.r.t. the vector space $\omega_{\hor}^1(\frac{P\times \go}{G})$. The affine structure of the smooth connections is mirrored by the existence of a Lie groupoid morphism between the Lie groupoids $\Hol_{\Lambda}^P(\Sigma)$ and $\Hol_{\Lambda^\prime}^P(\Sigma)$ for a fixed gauge groupoid $\fG$. This is studied in the next corollary. 

\begin{cor}\label{cor liealgebroid} Let $\gamma_A$ and $\gamma_{A^\prime}$ be two Lie algebroid connections on the transitive Lie algebroids $A(\Hol_\Lambda^P(\Sigma))$ and $A(\Hol^P_{\Lambda^\prime}(\Sigma))$ associated to the holonomy Lie groupoids $\Hol^P_\Lambda(\Sigma)$ and $\Hol^P_{\Lambda^\prime}(\Sigma)$.  
 
Then there exists a Lie algebroid morphism $\abroid^\prime$ between $ A(\Hol_\Lambda^P(\Sigma))$ and $ A(\Hol^P_{\Lambda^\prime}(\Sigma))$. Moreover there is a Lie groupoid morphism $\mor^\prime$ from $\Hol_\Lambda^P(\Sigma)$ to $\Hol^P_{\Lambda^\prime}(\Sigma)$.
\end{cor}
\begin{proofs}
Consider the Lie algebroid $A(\Hol_\Lambda^P(\Sigma))$
and 
\beqs A(\Hol_{\Lambda^\prime}^P(\Sigma)):= \left\{ X\in\frac{TP}{G}: X-(\gamma_{A^\prime}\circ\pi_*)(X)\in\Ima(\bar R_{A^\prime}) \right\}
\eqs
Recall that $\gamma_{A^\prime}:=\gamma_A + j\circ l$ holds.
Derive
\beqs  &j(\bar R_{A}(X,Y)) -j(\bar R_{A^\prime}(X,Y))= \gamma_{A}[X,Y]-[\gamma_{A}X,\gamma_{A}Y] -\gamma_{A^\prime}[X,Y] 
+ [\gamma_{A^\prime}X,\gamma_{A^\prime}Y] \\
&=\gamma_{A}[X,Y]-[\gamma_{A}X,\gamma_{A}Y] -\gamma_A [X,Y] + (j\circ l)[X,Y] 
+ [(\gamma_A + j\circ l)X,(\gamma_A + j\circ l)Y] \\
&=-[\gamma_{A}X,\gamma_{A}Y]  + (j\circ l)[X,Y] 
+ [\gamma_A X,\gamma_A Y] +  [(j\circ l)X, (j\circ l)Y] \\
&=(j\circ l)[X,Y] +  [(j\circ l)X, (j\circ l)Y] =: j(\bar R_{l}(X,Y))
\eqs
Hence obtain
\beqs A(\Hol^P_{\Lambda^\prime}(\Sigma)):= \left\{ X\in\frac{TP}{G}: X-(\gamma_{A}\circ\pi_*)(X)-((j\circ l)\circ\pi_*)(X)\in\Ima(\bar R_A+\bar R_l )\right\}
\eqs
Consequently for $X\in A(\Hol^P_\Lambda(\Sigma))$
\beqs \abroid_l(X)=\abroid_l((\gamma_{A}\circ\pi_*)(X))= (\gamma_{A}\circ\pi_*)(X) + ((j\circ l)\circ\pi_*)(X)=:\abroid^\prime(X)
\eqs defines a  Lie algebroid morphism $\abroid^\prime$ from $A(\Hol^P_{\Lambda}(\Sigma))$ to $A(\Hol^P_{\Lambda^\prime}(\Sigma))$ over the same base $\Sigma$. This is verified by proving that $\abroid_l$ is a vector bundle morphism such that the anchor preserving condition
\beq a^\prime =a\circ \abroid^\prime
\eq
and the bracket condition
\beq \abroid^\prime(\bra X,Y\ket)=\bra \abroid^\prime(X),\abroid^\prime(Y)\ket
\eq
are satisfied. Remember $j\circ\omega^{/G}+ \gamma_A\circ \pi_\ast=\id_{A\Goid}$ and consequently it is true that,
\beqs &\gamma_{A}\circ\pi_*+ (j\circ l)\circ\pi_*
= j\circ\omega +j\circ\omega^{\prime}
\eqs whenever $\omega,\omega^\prime\in  \Omega^1_{\basic}(P,\go)^G$ yields. Then derive $j\circ\omega +j\circ\omega^\prime=j\circ \tilde\omega\in\Omega_{\basic}^1(P,\go)^G$.

Finally define a morphsim $\mor^\prime: \Hol^P_{\Lambda}(\Sigma)\longrightarrow \Hol^P_{\Lambda^\prime}(\Sigma)$ by $\abroid^\prime= : \mor^\prime_*$ such that
\beqs \gamma_{A^\prime}= \mor^\prime_*\circ\gamma_A =\abroid_l\circ\gamma_A
\eqs and
\beqs \Lambda^\prime=\mor^\prime\circ\Lambda
\eqs is fulfilled.
Then $\mor^\prime$ is a Lie groupoid morphism.
\end{proofs}
Notice that, the failure of an infinitesimal Lie algebroid connection $\gamma_A$ to be a Lie algebroid morphism is given by the curvature, i.e.
\beqs j(\bar R_{A}(X,Y)) = \gamma_{A}[X,Y]-[\gamma_{A}X,\gamma_{A}Y] 
\eqs whenever $X,Y\in A(\Hol_\Lambda^P(\Sigma))$.

\begin{defi} Let $\breve L$ be the set of all maps $l:T\Sigma\longrightarrow \frac{P\times\go}{G}$ such that $\gamma_{A^\prime}=\gamma_A +j\circ l$ and $\gamma_A,\gamma_{A^\prime}$ are right splittings of the Atiyah sequence. Set $\Lambda^\prime:=\Lambda_l$.

Then the \textbf{set of holonomy maps for a gauge theory} is defined by
\beqs \Hol^P_{\breve\Lambda}(\Sigma):=\{ (\mor_l\circ\ho_\Lambda)(\gamma)  \in\Hol^P_{\Lambda_l}(\Sigma):\gamma\in\PD\Sigma, l\in\breve L,\Lambda\in\breve \Lambda \}
 \eqs  where $\Goid=\frac{P\times P}{G}$.
\end{defi}
This set replaces the configuration space $\breve \A$ of smooth connections for a gauge theory.
The set of holonomy maps for a gauge theory corresponds one-to-one to the set, which is given by
\beqs A(\Hol^P_{\breve\Lambda}(\Sigma)):= \left\{ X\in\frac{TP}{G}: X-(\gamma_{A}\circ\pi_*)(X)-((j\circ l)\circ\pi_*)(X)\in\Ima(\bar R_A+\bar R_l )\forall l\in\breve L\right\}
\eqs Summarising this set is derived from the Lie algebroids associated to holonomy Lie groupoids.
\section{Holonomy groupoid and holonomy-flux groupoid $C^*$-algebras for gauge theories}\label{sec holgroupoid}

In this section some ideas for a new construction of algebras in the holonomy groupoid formulation of LQG is presented. The aim is to find a suitable algebra such that the curvature is contained in this new algebra. In the section \ref{sec duality} it has been argued that, curvature and the infinitesimal connection have the same base, since their values are contained in the Lie algebroid of the holonomy groupoid. This is the new starting point of the construction of algebras in the next section \ref{subsec Calgebraholgroupoid}. The fundamental idea of Barrett has been to declare the set of all holonomy maps to be the configuration space of the theory. Consequently a set of algebras depending on holonomy groupoids associated to path connections generalises this choice. The implementation of the flux operators is indicated in section \ref{subsec Calgebracrossgroupoid} by a cross-product construction, which is similar to the definition of the holonomy-flux cross-product $C^*$-algebra \cite{Kaminski2}, \cite[Chapt.: 7]{KaminskiPHD}.
Furthermore there are morphisms between holonomy Lie groupoids associated to different path connections, which correspond to relations between Lie algebroids. These relations are used to define morphisms between algebras. Since the construction depends on the choice of the base manifold $\Sigma$ a new covariant formulation is suggested. The author proposes in section \ref{subsec background} to use for the covariant holonomy groupoid formulation of LQG the ideas, which have been presented by Brunetti, Fredenhagen and Verch \cite{BrunFredVerch01} in the context of algebraic quantum field theory.

\subsection{The construction of the holonomy groupoid $C^*$-algebra for gauge theories}\label{subsec Calgebraholgroupoid}
The ideas for a construction of algebras in the holonomy groupoid formulation is not straight forward. Therefore a couple of different approaches are analysed. Some of these are not available for LQG. 

The first idea is to use the theory of locally compact Hausdorff groupoids, which is influenced by the theory of locally compact groups. Indeed Renault \cite{Renault80} has presented $C^*$-algebras constructed from locally compact Hausdorff groupoids. The idea is the following. In comparison with the group algebra of a locally compact group, a similar groupoid algebra is constructed. The space of continuous functions with compact support on a groupoid, which is equipped with a convolution multiplication and an involution, form a $^*$-algebra. There is a $C^*$-norm such that the representations of that algebra are continuous.  In analogy to Haar measures on locally compact groups a system of Haar measures is derived. 
Now recall the holonomy groupoid $\Hol_\Lambda(\Sigma)$, which has been presented in section \ref{subsec pathconnection}, and the holonomy groupoid $\Hol_\Lambda^P(\Sigma)$ for a gauge theory, which has been given in section \ref{subsec holmapsgaugetheories2}. Then in general it is not clear, if the holonomy groupoid  $\Hol_\Lambda^P(\Sigma)$ is locally compact and Hausdorff. Consequently this approach by Renault cannot directly being used in this context.

The second idea is to consider the $^*$-algebra $\CD^*(\Goid)$ of continuous functions on the Lie groupoid $\Goid$, which is for example given by the gauge groupoid $\fG$. This algebra is the analog of the convolution $^*$-algebra $\CD^*(G)$ for a locally compact group $G$. Then the groupoid $C^*$-algebra $C^*(\Goid)$ is isomorphic to $\KD(L^2(P))\otimes C^*(G)$. This result has been stated by Landsmann \cite{Land}. But the $C^*$-algebra $C^*(\Goid)$ is not the right choice, since the particular holonomy groupoid structure is absent and the full knowledge of the manifold $P$, or the base manifold $\Sigma$ and a section $s: \Sigma\rightarrow P$, is needed. From this point of view, this algebra is maybe not the favoured algebra. Consequently one might use this idea for gravitational theories instead of a pure gauge theory.

The third idea is based on transitive Lie groupoids. Assume that the holonomy groupoid $\Hol_{\Lambda}^P(\Sigma)$ associated to a principal bundle $P(\Sigma,G,\pi)$ is a transitive Lie groupoid. Then another construction of a $C^*$-algebra is available. Notice that, $\Hol_{\Lambda}^P(\Sigma)$ is a Lie subgroupoid of the gauge groupoid $\fG$.
Then the holonomy groupoid $C^*$-algebra for a gauge theory is defined as follows.
Landsman \cite[Definition 3.3.2]{Landsman98} has defined a family of measures for a  Lie groupoid. 
\begin{defi}
A \textbf{left Haar system on a Lie groupoid} $\Group$ is a family $\{\mu_v^{t_{\Goid}}\}_{v\in\Goid^0}$ of positive measures, where $\mu_v^{t_{\Goid}}$ is a measure on the manifold $t_{\Goid}^{-1}(v)$ such that
\begin{enumerate}
 \item\label{en leftinv} the family is invariant under the \hyperlink{left-translation in a Lie groupoid}{left-translation in a Lie groupoid} $\Goid$
 \item each $\mu_v^{t_{\Goid} }$ is locally Lebesque (i.e. a Lebesque measure in every co-ordinate chart)
 \item for each $f\in \CD(\Goid)$ the map $v\mapsto \int_{t_{\Goid}^{-1}(v)}\dif \mu_v^{t_{\Goid}}(\gamma)f(\gamma)$ from $\Goid^0$ to $\CB$ is smooth.
\end{enumerate}
 \end{defi}
Equivalently a right Haar system on a Lie groupoid is defined. Consequently there is a left Haar measure on $\Hol^P_\Lambda(\Sigma)$. Set $\Goid:=\Hol_{\Lambda}^P(\Sigma)$.
Note that, for a Lie groupoid $\Goid$ over $\Goid^0$ the property \ref{en leftinv} induces
\beqs \int_{t_{\Goid}^{-1}(v)}\dif \mu_v^{t_{\Goid}}(\gamma)f(\gamma)
=\int_{t_{\Goid}^{-1}(v)}\dif \mu_v^{t_{\Goid}}(L_\theta(\gamma))f(L_\theta(\gamma))
\eqs for every $\theta\in\Goid^{\varphi_0(v)}_v$ and $(\varphi,\varphi_0)\in\Diff(\Goid)$.
Furthermore it is possible to consider the reduced groupoid $C^*$-algebra $C^*_r(\Goid)$, which is defined in analogy to the reduced group $C^*$-algebra of a Lie group. Clearly the algebra depends on the choice of the left Haar system on $\Group$.
The convolution product of two functions $f,k\in\CD(\Goid)$ is given by
\beqs (f \ast k)(\gamma):=\int_{t_{\Goid}^{-1}(s_{\Goid}(\gamma))}\dif\mu_{s_{\Goid}(\gamma)}^{t_{\Goid}}(\tg) f(\gamma\circ\tg)k(\tg^{-1})
\eqs
and involution is presented by $f^*(\gamma):=\overline{f(\gamma^{-1})}$.
There is a norm of $\CD(\Goid)$ defined by
\beqs\|f\|_2:=\int_{t_{\Goid}^{-1}(s_{\Goid}(\gamma))}\dif\mu_{s_{\Goid}(\gamma)}^{t_{\Goid}}(\gamma) \vert f(\gamma)\vert^2
\eqs such that the completition with respect to this norm defines the reduced groupoid $C^*$-algebra $C^*_r(\Goid)$. 
\begin{defi}
The reduced groupoid $C^*$-algebra for the holonomy Lie groupoid $\Hol^P_\Lambda(\Sigma)$ is called the \textbf{holonomy groupoid $C^*$-algebra for a gauge theory associated to a path connection $\Lambda$} and is denoted by $C^*(\Hol_{\Lambda}^P(\Sigma))$.
\end{defi}

Recall from section \ref{subsec Liegroupoids} that for every Lie groupoid $\Goid$ there exists an associated Lie algebroid $A\Goid$. This is a vector bundle over $\Goid^0$, which is equipped with a vector bundle map $\Goid\rightarrow T\Goid^0$, a Lie bracket $\bra.,.\ket_{\Goid}$ on the space $\Gamma(\Goid)$ of smooth sections of $\Goid$, satisfying certain compatibility conditions. 
Similarly to the exponentiated map from a Lie algebra $\go$ associated to a Lie group $G$ to $G$ a generalised exponentiated map has been mentioned in section \ref{sec duality}. 

Now remember it has been assumed that, $\Hol^P_{\Lambda}(\Sigma)$ defines a holonomy Lie groupoid for each path connection $\Lambda$ in $\breve \Lambda$. There exists an associated Lie algebroid $\A\Hol^P_{\Lambda}(\Sigma)$ for each path connection $\Lambda$ in $\breve \Lambda$. This Lie algebroid contains the values of the infinitesimal Lie algebroid connections and the curvature. Moreover a Lie algebroid morphism $\abroid_l$ between two Lie algebroids and the groupoid morphism $\mor_l$ between two holonomy Lie groupoids associated to different path connections have been presented in corollary \ref{cor liealgebroid}.
Let $\Lambda$ and $\Lambda_l$ be two path connections in $\breve \Lambda$. 
Then there is a morphism, which depends on the Lie algebroid morphism $\abroid_l$, between the reduced groupoid $C^*$-algebras. 
\begin{defi}Let $\abroid_l$ be a Lie algebroid morphism between $\Hol^P_{\Lambda}(\Sigma)$ and $\Hol^P_{\Lambda_l}(\Sigma)$.

There is a morphism $\alpha_{\abroid_l}:C^*(\Hol^P_{\Lambda}(\Sigma))\rightarrow C^*(\Hol^P_{\Lambda_l}(\Sigma))$ defined by
\beq (\alpha_{\abroid_l} f)(\ho_\Lambda(\gamma))
:=f^l((\mor_l\circ\ho_\Lambda)(\gamma))=f^l(\ho_{\Lambda^\prime}(\gamma))
\eq whenever $f\in C^*(\Hol^P_{\Lambda}(\Sigma))$ and $f^l\in C^*(\Hol^P_{\Lambda_l}(\Sigma))$.
\end{defi}

The left generalised exponentiated map $\Exp_L:\Gamma A(\Hol^P_{\Lambda}(\Sigma))\rightarrow \Gamma\Hol^P_{\Lambda}(\Sigma)$, 
which is defined for all right-invariant vector fields in a vector subspace of $\frac{TP}{G}$, leads to an action on $C^*(\Hol^P_{\Lambda}(\Sigma))$. 
\begin{defi}Let $\gamma_A$ be a Lie algebroid connection associated to the path connection $\Lambda$.

There is an action $\alpha_{\gamma_A}:C^*(\Hol^P_{\Lambda}(\Sigma))\rightarrow C^*(\Hol^P_{\Lambda}(\Sigma))$ defined by
\beqs (\alpha_{\gamma_A}f)(\ho_\Lambda(\gamma)) 
:=f(L_{\Exp_L(t (\gamma_A(X))(v))}(\ho_\Lambda(\gamma)))
=f(\Exp_L(t (\gamma_A(X))(v))\ho_\Lambda(\gamma))
\eqs where $t\in\R$, $X\in T_v\Sigma$ and $X:=\frac{\dif}{\dif t}\Big\vert_{t=t_0}\gamma(t)$, $v=t(\gamma)$ such that $\Exp_L(t(\gamma_A(X))(v))\in\Hol^P_{\Lambda}(\Sigma)^v$ and $f\in C^*(\Hol^P_{\Lambda}(\Sigma))$.
\end{defi}

Similarly an action of a bisection $\sigma_A$  of $\fG$ is given.
It is true that, $\Exp(\gamma_A(X)(v))=\Lambda(\varphi,v)(1)$ holds and $\tilde\varphi_t(\phi_t,\id_G)$ defines a gauge and diffeomorphism transformation on $\fG$. 
\begin{defi}Let $\sigma_A$ be a bisection of the gauge groupoid $\fG$. 

There is an action $\alpha_{\sigma_A}:C^*(\Hol^P_{\Lambda}(\Sigma))\rightarrow C^*(\Hol^P_{\Lambda}(\Sigma))$ defined by
\beqs (\alpha_{\sigma_A} f)(\ho_\Lambda(\gamma)):= f(L_{\tilde\varphi(\phi,\id_G)}\ho_\Lambda(\gamma)) :=f((\id_\Lambda\circ\ho_\Lambda)(\varphi\circ\gamma))
=f(\ho_\Lambda(\varphi)\ho_\Lambda(\gamma))
\eqs where $\phi(v)=s(\varphi)=:w$, $t(\gamma)=k$, $\gamma\in\PD\Sigma^v_k$, $\varphi\in\PD\Sigma^w_v$,  $\id_\Lambda: \Hol^P_\Lambda(\Sigma)\longrightarrow \Hol^P_\Lambda(\Sigma)$ is the identity morphism and $f\in C^*(\Hol^P_{\Lambda}(\Sigma))$. 
\end{defi}
This action is generalised to an action of a bisection $\sigma_{A^\prime}$  of $\fG$: 
\beqs (\alpha_{\sigma_{A^\prime}} f)(\ho_\Lambda(\gamma)):= f(L_{\tilde\varphi^\prime(\phi,\id_G)}\ho_\Lambda(\gamma)) :=f((\mor_l\circ\ho_\Lambda)(\varphi\circ\gamma))=f(\ho_{\Lambda^\prime}(\varphi\circ\gamma))
\eqs whenever $\Lambda^\prime=\mor_l\circ\Lambda$, $\abroid_l:= \abroid^\prime$, $f\in C^*(\Hol^P_{\Lambda}(\Sigma))$ and $\gamma_{A^\prime}=\mor^\prime_*\circ\gamma_A$.

\subsection{Cross-product $C^*$-algebras for gauge theories}\label{subsec Calgebracrossgroupoid}

In the last section the holonomy groupoid $C^*$-algebra for a gauge theory associated to a path connection has been presented. This algebra is the algebra of quantum configuration variables. The full algebra is derived in an analogue procedure presented in \cite{Kaminski2,KaminskiPHD}. Indeed Masuda  \cite{MasudaII} has invented cross-product $C^*$-algebras in the context of groupoids. The aim is to use these cross-product construction in the holonomy groupoid formulation. 
\begin{defi}
The triplet $(\Alg,\Goid,\rho)$ is called a \textbf{$C^*$-groupoid dynamical system} iff $\Alg$ is a $C^*$-algebra, $\Goid$ is a locally compact groupoid with a faithful transverse function $\mu=\{\mu^v\}_{v\in\Goid^0}$ and $\rho:\Goid\longrightarrow \Aut(\Alg)$ is a continuous morphism.
\end{defi}

First it has been used in \cite{Kaminski2,KaminskiPHD} that, the quantum flux operators are implemented as elements of a Lie group $G$.  
Concern the holonomy Lie groupoid is $\Goid:=\Hol^P_{\Lambda}(\Sigma)$. Then the following $C^*$-groupoid dynamical system $(C_0(G),\Goid,\rho_L)$, where $\rho_L: \Goid\longrightarrow G$ is a continuous groupoid morphism, is studied. The associated cross-product $C_0(G)\rtimes_L\Goid$ is defined as the completition of the set $\CD(\Goid,\Alg)$ of all $\Alg$-valued continuous functions over $\Goid$ with compact support with respect to a appropriate $C^*$-norm. Clearly there are left actions $\rho_L$ and right actions $\rho_R$ of $\Goid$ on the algebra $C_0(G)$ associated to left or right Haar systems  on the holonomy Lie groupoid $\Goid$. 

Consequently the following algebra contains holonomies and flux operators for a gauge theory.
\begin{defi}
The \textbf{holonomy-flux groupoid $C^*$-algebra for a gauge theory associated to a path connection $\Lambda$} is defined by the cross-products $C_0(G)\rtimes_{L}\Hol^P_{\Lambda}(\Sigma)$ or $C_0(G)\rtimes_{R}\Hol^P_{\Lambda}(\Sigma)$.
\end{defi}
Summarising these algebras are the algebras of quantum configuration and quantum momentum variables for a gauge theory. But recognize that there is a big bunch of these algebras, since each algebra is associated to a path connection. Moreover each algebra really depends on the chosen gauge theory and hence on the principal fibre bundle $P(\Sigma,G,\pi)$. This leads directly to sets of algebras. 

\subsection{Covariant holonomy groupoid formulation of LQG}\label{subsec background}

In this section two categories, which arise naturally in the holonomy groupoid formulation of LQG are presented.
First recall the set $\Hol^P_{\breve\Lambda}(\Sigma)$ of holonomy Lie groupoids, which has been analysed in section \ref{subsubsec holmapdef}. Then one category is given by the following objects and morphisms. The set of objects is formed by all holonomy groupoids $\Hol_{\Lambda}(\Sigma)$ associated to each manifold $\Sigma$ in a set $\breve\Sigma$ of $3$-dimensional spatial manifolds and to each path connection $\Lambda$ in a set $\breve\Lambda$ of path connections w.r.t. a principal fibre-bundle $P(\Sigma,G_\Sigma,\pi)$. Notice that the structure group $G_\Sigma$ vary for principal fibre bundles associated to different base manifolds. The set of morphisms of the category are Lie groupoid morphism between two holonomy Lie groupoids. Denote this category by $\mathfrak{Hol}$ and call it the \textbf{holonomy category}. 

The second category is given by the objects, which are given by all unital $C^*$-algebras $C_0(G_\Sigma)\rtimes_{X}\Hol_{\Lambda}(\Sigma)$ for every principal fibre bundle $P(\Sigma,G_\Sigma,\pi)$ containing a manifold $\Sigma$ in $\breve \Sigma$ and a Lie group $G_\Sigma$ associated to $\Sigma$, and defined by the left and right actions ($X=R,L$).  The set of morphisms are faithful unit-preserving $^*$-morphisms between these algebras. The category is denoted by $\mathfrak{Alg}$ and called the \textbf{holonomy-flux category}. 

Summarising a covariant holonomy groupoid formulation of LQG is an assignment of $C^*$-algebras to holonomy groupoids in such a way that the algebras are identifyable if the holonomy Lie groupoids are connected by a Lie groupoid morphism.    

In this article an overview about a new formulation of LQG has been given. This is a starting point for a detailed study, which will be done in a future work.

\section*{Acknowledgements}
The work has been supported by the Emmy-Noether-Programm (grant FL 622/1-1) of the Deutsche Forschungsgemeinschaft.

\end{document}